\documentclass[10pt,conference,twocolumn,letterpaper]{IEEEtran}

\usepackage[latin1]{inputenc}
\usepackage[american]{babel}
\usepackage{cite}
\usepackage{soul}
\usepackage[disable]{todonotes} %put [disable] if needed
\usepackage{algorithmic}
\usepackage{array}
\usepackage{fixltx2e}
\usepackage{graphicx}
\usepackage{booktabs}
\usepackage{url}
\usepackage{xspace}
\usepackage{multirow}
\usepackage[normalem]{ulem}
\usepackage{ntheorem}
\usepackage{amssymb}
\usepackage{amsfonts}
\usepackage{amsmath}
\usepackage[numbers,sort,square]{natbib}
\usepackage{bm}
\usepackage{relsize}

\newcommand{\thesystemtitle}{Phoenix}
\newcommand{\thesystem}{\textsf{\relsize{-.5}{\thesystemtitle}}\xspace}

\renewcommand{\hl}{}

\newcommand{\subparagraph}{}
\usepackage{titlesec}
\titleformat{\paragraph}[runin]{\normalfont\normalsize\bfseries}{}{0pt}{}[.]
\titlespacing*{\paragraph}{0pt}{8pt}{0.5em}

\DeclareMathOperator{\IPs}{IPs}

\DeclareMathOperator{\dmahop}{d}
\newcommand{\dmah}[1]{\dmahop_{Mah}(#1)}

\begin{document}

%\title{Characterizing Automatically Generated Domains}
\title{Tracking and Characterizing Botnets Using Automatically
Generated Domains}
%\author{Stefano Schiavoni \and Federico Maggi \and Lorenzo Cavallaro
%and Stefano Zanero}

\author{\IEEEauthorblockN{Stefano Schiavoni}
\IEEEauthorblockA{
Politecnico di Milano}
\and
\IEEEauthorblockN{Federico Maggi}
\IEEEauthorblockA{Politecnico di
Milano}
\and
\IEEEauthorblockN{Lorenzo Cavallaro}
\IEEEauthorblockA{Royal Holloway University of
London}
\and
\IEEEauthorblockN{Stefano Zanero}
\IEEEauthorblockA{Politecnico di
Milano}}

\maketitle

\begin{abstract}
  Modern botnets rely on domain-generation algorithms (DGAs) to build
  resilient command-and-control infrastructures. Recent works focus
  on recognizing automatically generated domains (AGDs) from DNS
  traffic, which potentially allows to identify previously unknown
  AGDs to hinder or disrupt botnets' communication capabilities.

  The state-of-the-art approaches require to deploy low-level DNS
  sensors to access data whose collection poses practical and privacy
  issues, making their adoption problematic. We propose a mechanism
  that overcomes the above limitations by analyzing DNS traffic data
  through a combination of linguistic and IP-based features of
  suspicious domains. In this way, we are able to identify AGD names,
  characterize their DGAs and isolate logical groups of domains that
  represent the respective botnets. Moreover, our system enriches
  these groups with new, previously unknown AGD names, and produce
  novel knowledge about the evolving behavior of each tracked botnet.

  We used our system in real-world settings, to help researchers that
  requested intelligence on suspicious domains and were able to label
  them as belonging to the correct botnet automatically.

  Additionally, we ran an evaluation on 1,153,516 domains, including
  AGDs from both modern (e.g., Bamital) and traditional (e.g.,
  Conficker, Torpig) botnets. Our approach correctly isolated families
  of AGDs that belonged to distinct DGAs, and set automatically
  generated from non-automatically generated domains apart in 94.8
  percent of the cases.
\end{abstract}

\section{Introduction}
\label{sec:introduction}
Botnets continue to play a significant role in today's cybercrime
ecosystem, and have become a commodity platform for online lucrative
activities. The latest ENISA Threat Landscape~\cite{Marinos:2012ww}
highlighted that one of the adverse effects of this
malware-as-a-service trend is an increased number of small, distinct
botnets, which are predicted to replace traditional large botnets:
Smaller botnets are likely to fly under the radar because of their
size. This creates the problem of keeping track of such a diverse
population of distinct, yet related threats. Furthermore, botnets are
used for a variety of malicious purposes, ranging from spamming to
information stealing and espionage.

Identifying malicious activities related to botnets is a well-studied
problem with many proposed solutions. One of the most promising
directions consists in observing and obstructing the traffic of their
communication infrastructure. Typically, the security defenders strive
to find the IPs or the domain names of the command-and-control (C\&C)
server of a botnet with the goal of creating sinkholing IPs:
Eventually, the bots will starve not being able to contact their
masters.

The most reliable yet easy-to-manage bot-to-master centralized
communication mechanism relies on domain
flux~\cite{Stone-Gross:2009:YBM:1653662.1653738}: Time-dependent
rendezvous points via domain-generation algorithms (DGAs). Each
bot of a given botnet is shipped with the same DGA, which
generates a large number of time-dependent C\&C domain names the bots
will contact over time.
% Here, the bots and the master agree on pseudo-random algorithms that
% generate a large number of domain names where the bot tries to
% contact its master.
The key concept is that, at any moment, only a small number of domains
is active and resolves to the true IPs of the C\&C server. This
characteristic makes dealing with DGAs very expensive for the security
defenders, because they would need to track---and register, in case of
botnet sinkholing attempts---several thousands of domains before finding the
one used as the rendezvous point.

\paragraph{Research Gaps}
Researchers have proposed various approaches for finding automatically
generated domains (AGDs) with the goal of correlating this information
with other datasets (e.g., blacklists or domain reputation scores). We
notice two main shortcomings. On the one hand, the existing approaches
provide rich details on \emph{individual} malicious domains
(e.g.,~\citet{bilge2011exposure,antonakakis2011detecting}). However,
they fail in \emph{correlating} distinct but related abuses of such
domains. On the other hand, the approaches that are able to
automatically correlate malicious domains to the botnets that lie
behind them (e.g.,~\citet{antonakakisthrow}) require access to
Internet traffic whose collection poses \emph{practical} and
\emph{privacy} issues. More precisely, the approach
by~\citet{antonakakisthrow} requires visibility on the infected
machines' IP address. Besides privacy issues, this creates the
research problem of repeatability and, more importantly, requires a
low-level traffic sensor deployed between the infected machines and
the DNS servers that they contact, with visibility of the original DNS
queries. Moreover, the accuracy of this approach may be affected by
IP-sharing (e.g., NAT) or IP-reusing (e.g., DHCP)
mechanisms~\cite{Stone-Gross:2009:YBM:1653662.1653738}.

\paragraph{Proposed Approach}
We propose \thesystem, which leverages publicly available passive DNS
traffic to (1) find AGDs, (2) characterize the generation algorithms,
(3) isolate logical groups of domains that represent the respective
botnets, and (4) produce novel knowledge about the evolving behavior
of each tracked botnet. \thesystem requires no knowledge of the DGAs
active in the wild and, in particular, no reverse engineering of the
malware. Being based on recursive-level passive DNS traffic, our
approach guarantees scientific repeatability of malware
experiments~\cite{rossow2012prudent}, preserves the privacy of the
infected computers, and is not affected by any IP-sharing or
IP-reusing mechanisms in place. 

\thesystem \emph{uses} the output of previous work that identify
malicious domains from passive DNS monitoring; it processes these
feeds of malicious domains to recognize the typical patterns of AGDs
and build knowledge and insights about their provenance to help botnet
investigations.
First, \thesystem creates linguistic models of \emph{non}-AGDs (e.g.,
benign domains). Domains that violate such models are considered to be
automatically generated. Then, \thesystem groups these domains
according to the domain-to-IP relations. From these groups, \thesystem
derives a generic set of fingerprints to label new domains as
belonging to some botnet. Such fingerprints are useful to characterize
the evolution of botnets, gather insights on their activity (e.g.,
migrations), and identify previously unknown C\&C domain names (e.g.,
new blacklists).

\paragraph{Original Contributions} Our approach
\begin{itemize}
\item identifies groups of AGDs and models the characteristics of the
  generation algorithms, with less requirements than previous work;
  with this we ensure a) privacy-preserving data collection, b) ease
  of large-scale deployment, c) repeatable evaluation, and d)
  resiliency from accuracy-affecting IP-sharing (e.g., NAT) and
  IP-reusing (e.g., DHCP) mechanisms;
\item automatically associates \emph{new} malicious domains to the
  activity of botnets;
\item builds \emph{new correlated knowledge} that allow security
  analysts to track the evolution of a DGA-based botnets.
\end{itemize}

\paragraph{Results}
We evaluated \thesystem on 1,153,516 real-world
domains. \thesystem correctly isolated families of domains that belong
to different DGAs. Also, on February 9th we obtained an undisclosed
list of AGDs for which no knowledge of the respective botnet was
available. \thesystem labeled these domains as belonging to Conficker:
Further investigation eventually confirmed that it was indeed
Conficker.B. We believe that, in addition to the comprehensive
evaluation, this latter fact proves the practicality and effectiveness
of our approach.

\section{AGD Names}
\label{sec:motivation}
Thanks to the Domain Name System (DNS), applications and users do not
need to keep track of IP addresses, but can use human-readable
aliases, called humanly generated domain (HGD) names from
hereinafter. The hierarchical structure of the DNS protocol
efficiently maintains cached copies of the name-IP associations, so
that the resolution is fast, reliable, and highly distributed. The
domain name resolution has many uses, from load balancing to failover
setups. Unfortunately, these characteristics are useful for malicious
developers, too.  Miscreants began using DNS to make
\emph{centralized}\footnote{Although DNS may be used by botnets of
  arbitrary network topology, in this paper we focus on centralized
  domain flux botnet infrastructures.} botnet infrastructures more
reliable and, more importantly, harder to map and take
down~\cite{antonakakisthrow,perdisci2012early,antonakakis2011detecting,passerini2008fluxor,holz2008measure,yadav2012detecting,Stone-Gross:2009:YBM:1653662.1653738}. In
particular, botnets rely on the DNS (also) to implement call-home
functionalities, such that the bots can contact their C\&C server to
receive instructions, updates, or to upload stolen data. For instance,
the botnet operator, commonly referred to as the \emph{botmaster}, can
change the IP address of the C\&C dynamically, by updating the DNS
record, for instance when he or she needs to migrate the C\&C to other
(more lenient) autonomous systems (ASs). Similarly, the botmaster can
associate short-lived multiple IPs of infected hosts to an individual
DNS domain over time~\cite{holz2008measure,passerini2008fluxor} to
improve the botnet's redundancy, while hiding the IP-based coordinates
of a malicious host (i.e., the mothership).

In this context, a very effective technique to improve centralized
botnets' resiliency to take downs and tracking is domain
flux~\cite{Stone-Gross:2009:YBM:1653662.1653738}, which resorts to
equip a bot with DGAs.

\paragraph{Domain-generation Algorithms} DGAs have become the
technique of
choice\footnote{\url{https://blog.damballa.com/archives/1906}} for
building effective rendezvous mechanisms between bots and their C\&C
servers: The bots of a given botnet implement the same algorithm to
generate a large and time-dependent list of AGDs, based on
pseudo-unpredictable seeds (e.g., trending topics on Twitter, results
of a Google search). Only one of these AGDs is actually registered and
points to the true IP address of the C\&C. The bots will then
generate and query all these domains, according to the DGA, until a
DNS server answers with a non-NXDOMAIN reply, that is the IP address
of the respective (existing) AGD. The key is that only the DGA authors
(or the botmasters) know exactly when the upcoming rendezvous domain
has to be registered and activated\footnote{Differently from current
  DGAs, former forms of DGA (e.g., those based on sequential queries
  of AGDs) and failure to promptly register domains have allowed
  botnet takeovers in the
  past~\cite{Stone-Gross:2009:YBM:1653662.1653738}.}.

Finding ``families'', or groups, of related AGDs is therefore
fundamental, because they provide valuable insights to recognize, for
instance, sets of botnets that implement a DGA with common
characteristics---which refer, possibly, to the same botnet, or an
evolution. Therefore, the analysts can follow the evolution of these
botnets and their (changing) C\&Cs over time, where these are hosted,
and the number of machines involved. The task of finding families of
related AGDs, however, is tedious and labor-intensive, although
previous research has devised mechanisms to partially automate it.

In this work, we go beyond the state of the art and observe that
instances of AGD names drawn from the same DGA can be generalized into
``fingerprint'' of the generation algorithm itself, without reversing
its implementation. This boosts the detection and classification of
new, previously unknown, AGD names, thus enriching existing blacklists
with novel knowledge.

\paragraph*{Recognizing DGAs} A side effect of the DGA mechanisms
described above is that the bots will perform a disproportionately
large amount of DNS queries, which will result in NXDOMAIN
replies. This happens because the vast majority of such queries will
hit a non-existing domain. On the other hand, legitimate hosts have no
reasons to generate high volumes of queries that yield NXDOMAIN
replies. This observation has been leveraged
by~\citet{antonakakisthrow}, who proposed to detect DGA-based bots by
correlating the clients' IP found in the requests and the
corresponding NXDOMAIN replies at one or more DNS resolvers. This
allows to identify groups of bots while they are attempting to find
the rendezvous domain to reach their C\&C server. Also, the authors
show that their system finds families of AGDs with similar
characteristics (e.g., same upper-level domain name, lexicon or
syntax), possibly coined by the same DGA.

However, the NXDOMAIN responses alone carry little information that
can be directly used to identify the families of AGDs. Specifically,
an NXDOMAIN response only holds the queried domain name plus some
timing data. Therefore, as also noticed by~\citet{perdisci2012early},
the NXDOMAIN criterion requires knowledge of the querying hosts (i.e.,
the bots). For this reason, \citet{antonakakisthrow} had to rely on
the client IP to group together NXDOMAINS queried by the same set of
hosts. \citet{yadav2012winning} instead group together DNS queries
originated by the same client to define the correlation between
distinct requests that target the same domains.

\paragraph*{Drawbacks of Current Approaches} Setting aside the problem
of accuracy, which really matters for systems that tell malicious
vs. benign domains apart, we notice that current approaches impose
strict requirements in terms of input data required. Indeed, relying
on the IP addresses of the querying hosts has some drawbacks. First,
it is error prone, because no assumptions can be made on and
IP-(re)assignment and masquerading policies employed by ASs. Secondly,
and more importantly, obtaining access to this information is
difficult because it is available from DNS servers placed \emph{below}
the recursive DNS level (e.g., host DNSs). This can be problematic for
researchers, but also for practitioners who want to operate these
systems. Last, the need for information about the querying hosts
raises privacy issues and leads to non-repeatable
experiments~\cite{rossow2012prudent}, as datasets that include these
details are not publicly available. Moreover, the requirement
constrains the deployment of detection tools to the lowest level of
DNS hierarchy, preventing a large-scale, centralized use of the
proposed defending solutions.

\section{Goal and Challenges} After considering the above motivations,
we conclude that \emph{a method to find families of AGDs yielded by
  the same DGAs that requires no access to low- or top-level DNS data}
is necessary.

\paragraph{DGA Modeling Challenges} AGDs appear completely random at
sight. Nevertheless, creating automated procedures capable of modeling
and characterizing such ``randomness'' is hard, in particular when
observing one domain at a time, because one sample is not
representative of the whole random generation process. Grouping AGD
samples to extract the characteristics of the DGA is also challenging:
How to group AGDs samples together? How to avoid spurious samples that
would bias the results?

\paragraph{Data Collection Challenges} Collecting and dealing with DNS
traffic presents some challenges on (1) where to place the observation
point and (2) how to process the collected data. The amount of
information that can be collected varies depending on where the
sensors are deployed in the DNS hierarchy (e.g., low-level sensors
have access to the querying hosts, but are difficult to deploy,
whereas higher-level sensors are easier to deploy but their visibility
is limited to the stream of queried domains and their IPs). Moreover,
the domain-to-IP relations are highly volatile by their nature, with
both dense and sparse connections; this impacts the space and
computation complexity of the algorithms that are needed to analyze
DNS traffic.

\paragraph{Lack of Ground Truth} Last, little or no ground truth is
usually available regarding DGAs. When available, such ground truth is
limited to the knowledge of a specific implementation of DGA as a
result of a specific reverse engineering
effort~\cite{Stone-Gross:2009:YBM:1653662.1653738}. Therefore, this
knowledge is outdated once released and, more importantly, not
representative of the whole domain generation process but only of a
limited view.

\begin{figure*}
  % Source: https://docs.google.com/file/d/0BzyHJ8drzNDmbjVmLVotSVdxejA/edit
  \centering
  \includegraphics[width=.95\textwidth]{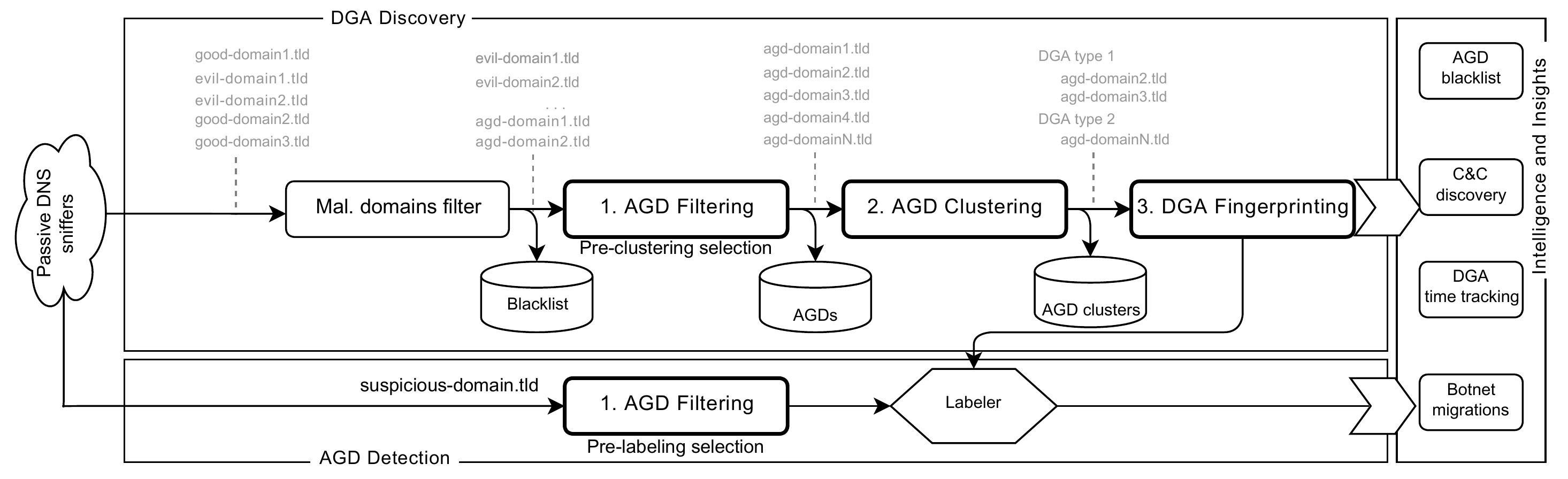}
  \caption{System overview of \thesystem. The \textbf{DGA Discovery}
    module processes the domain names from a domain reputation system
    and identifies interesting AGDs. The \textbf{AGD Detection} module
    analyzes a stream of DNS traffic and recognizes the names that
    appear to be automatically generated. This module also labels
    newly discovered AGDs as results of a particular DGA. The last
    module, \textbf{Intelligence and Insights}, provides the analyst
    with information useful, for instance, to track a botnet over
    time, as described in \S\ref{sec:intellicence-eval}.}
  \label{fig:system-overview}
\end{figure*}

\section{System Overview}
\label{sec:overview}

The following section provides a high-level overview of
\thesystem. All the system details are instead provided
in~\S{}~\ref{sec:details}.

\thesystem is divided into three modules (see
Fig.~\ref{fig:system-overview}). The core of \thesystem is the
\textbf{DGA Discovery} module, which identifies and models AGDs by
mining a stream of domains. The \textbf{AGD Detection} module receives
one or more domain names with the corresponding DNS traffic, and uses
the models built by the \textbf{DGA Discovery} module to tell whether
such domain names appear to be automatically generated. If that is the
case, this module labels those domains with an indication of the DGA
that is most likely behind the domain generation process. The
\textbf{Intelligence and Insights} module aggregates, correlates and
monitors the results of the other modules to extract meaningful
insights and intelligence information from the observed data (e.g.,
whether an unknown DGA-based botnet is migrating across ASs, whether a
previously unseen domain belongs to a particular AGD family).

\subsection{DGA Discovery Module}
\label{dga-discovery}
This module receives as input (1) a stream of domain names that are
known to be malicious and (2) a stream of DNS traffic (i.e., queries
and replies) \hl{collected above the recursive resolvers and} related
to such domains. This information is publicly accessible and can be
obtained easily from (1) a blacklist or domain reputation system
\hl{(e.g., \textsf{\relsize{-.5}{Exposure}}~\cite{bilge2011exposure})}
and (2) a passive \hl{and privacy-preserving} DNS monitor \hl{(e.g.,
  ISC/SIE~\cite{isc_sie})}. \hl{The blacklists that we rely on are
  generated from privacy-preserving DNS traffic too.} We make no
assumptions on the type of malicious activity  these domains are
used for (e.g., phishing websites, spamming campaigns or drive-by
download websites, botnet communication protocols).

\medskip\noindent This module follows a 3-step pipeline to recognize
domains that are used in DGA-based botnets.

\paragraph{Step~1 (AGD Filtering)} We extract a set of
\emph{linguistic features} from the domain names. The goal is to
recognize the (domain) names that appear to be the results of
automatic generations. For instance, we distinguish between
\texttt{5ybdiv.\hl{cn}}, which appears to be an AGD, and
\texttt{\hl{searchsmart.tk}}, which appears to be a HGD. For ease of
explanation and implementation, \thesystem considers the linguistic
features based on the English language, as discussed in
\S\ref{sec:discussion}.

Differently from previous work, we devised our features to work well
on single domains. Antonakakis et al.~\cite{antonakakisthrow} and
Yadav et al.~\cite{yadav2010detecting,yadav2012detecting}, instead,
rely on features extracted from \emph{groups} of domains, which raises
the issue on how to select such groups. The authors circumvented this
problem by choosing \emph{random} groups of domains. However, there is
no rigorous way to verify the validity of such an
assumption. Therefore, as part of our contributions, we made an effort
to design features that require no groupings of domains. We make no
assumptions about the type of DGA underlying the generated domains,
although we do assume that at least one exists.

The output is a set of AGDs, possibly generated by different
DGAs. \hl{We accomplish this using a \emph{semi-supervised
    techniques}, which only requires limited knowledge on HGDs (not on
  existing AGDs).}

\paragraph{Step~2 (AGD Clustering)} We extract a number of
\emph{IP-based features} from the DNS traffic of the domains that have
passed \textbf{Step~1}. We use these features to \emph{cluster
  together the AGDs} that have resolved to similar sets of IP
addresses---possibly, the C\&C servers. These \emph{AGD clusters} will
partition the AGDs according to the DNS replies observed. For example,
if the AGDs \texttt{5ybdiv.cn} and \texttt{hy093.cn} resolved to the
same pool of IPs, we will cluster them together. Here, we assume that
AGDs generated by different DGAs are used by distinct botnets, or
distinct malware variants, or at least by different botmasters, who
have customized or tuned a DGA for their C\&C strategy. Therefore,
this partitioning will, to some extent, reflect the different groups
of botnets that they have been employed for.

\paragraph{Step~3 (DGA Fingerprinting)} We extract a number of
features from the AGD clusters to create models that define the
fingerprints of the respective DGAs. It is worth noting that the set
of these features \emph{differs} from the one identified in the
previous step, as explained in detail in the next section.  The
\textbf{AGD Detection} module uses these fingerprints as a lookup
index to identify the type of \emph{previously unseen domains}. For
instance, as clarified in the remainder of this paper,
\texttt{epu.org} and \texttt{xmsyt.cn} will match two distinct
fingerprints.

%The output of this module is thus a set of clusters with their fingerprints.

% We assume that the AGDs within the same cluster will have the same
% characteristics, and thus the same \emph{type}.

\subsection{AGD Detection Module}
\label{agd-detection}
This module receives in input a (previously unseen) domain name $d$,
which can be either malicious or benign, and uses once again the
\textbf{AGD Filtering} step to verify whether it is automatically
generated.

The domain names that will pass this filter will undergo further
checks, which may eventually flag them as not belonging to any AGD
cluster (i.e., not matching any of the fingerprints). Therefore, in
this step, flagging as AGD a (benign) domain that do not belong to
some DGA is not a major error. It is instead more important not to
discard suspicious domains. Therefore, for this module only, we
configure the \textbf{AGD Filtering} step with a looser threshold (as
described in \S\ref{sec:agd-filtering}), such that we do not discard
domains that exhibit, even slightly, the linguistic characteristics
that are typical of AGDs. In other words, recall is more important
than precision in this module.

Then, this module leverages the AGD clusters and their respective
fingerprints to find the DGA, if any, that may lie behind the previously
unseen domain, $d$.

\subsection{Intelligence and Insights Module} Once an AGD is labeled
as such, it can be recognized among others as belonging to a given
``type'' of DGA. Therefore, the outcome of the previous modules
allows the extraction of summarized information and novel knowledge.

With this knowledge, the addresses of the C\&C servers and lists of
malicious AGDs can be grouped together in small sets that are easier
to analyze than if considered into one, large set, or distinctly. For
instance, if an analyst knows that 100 domains are malicious, he or
she can use the label information to split them into two smaller sets:
one that contains domain names ``similar'' to \texttt{5ybdiv.cn} and
\texttt{hy093.cn}, and one with domains ``similar'' to
\texttt{epu.org}.

The notion of ``similarity'' is by no means based solely on linguistic
similarity: We do consider other IP- and DNS-based features, explained
next. The top level domain is also not distinctive, we use it here as
a mere example. With this information, the analyst can track
separately the evolution of the IPs that the two groups point to and
take actions. For example, recognizing when a C\&C is migrated to a
new AS or undergoing a takedown operation is easier when the set of
IPs and domains is small and the characteristics of the DGA are known
and uniform.

Generally speaking, these analyses, for which we provided two use
cases in \S\ref{sec:intellicence-eval}, can lead to high-level
intelligence observations and conjectures, useful for the mitigation
of DGA-related threats. In this, we advance the state of the art by
providing a tool that goes beyond blacklists and reputation systems.

\section{System Details}
\label{sec:details}
We implemented \thesystem in \textsf{\relsize{-.5}{Python}} using the \textsf{\relsize{-.5}{NumPy}}
package, for statistical functions, and the
\textsf{\relsize{-.5}{SciPy}}~\cite{scipy} package, for handling sparse
matrices. Instead of relying entirely on existing libraries, we
implemented and customized the machine-learning algorithms described
in the remainder of this section for efficiency reasons. For this, we
leveraged domain-specific knowledge (e.g., data dimensionality).

\paragraph{Notation (Domain Names and Suffixes)}
For the purpose of this work, a \emph{domain name} is a sequence of
labels separated by dots (e.g.,
\texttt{www.example.com}). Specifically, the latter parts form the
\emph{public suffix} (e.g., \texttt{.com}, \texttt{.co.uk}), under
which Internet users can register a name, which is called \emph{chosen
  prefix} (e.g., \texttt{example}). The public suffix, which is often
referred to as top-level domain (TLD), can contain more than one label
(e.g., \texttt{.co.uk}). The term effective TLD (eTDL) is thus more
correct. eTLDs are enumerated at \url{http://publicsuffix.org/} to
allow parsing. A domain name can be organized hierarchically into more
subdomains (e.g., \texttt{www.example.com},
\texttt{ssh.example.com}). We only consider the first level of a
chosen prefix\hl{, simply because a DGA that works on further levels
  makes little sense, as the first level would still be the single
  point of failure}.

\subsection{Step~1: AGD Filtering}
\label{sec:agd-filtering}
In this step we make the assumption that if a domain is automatically
generated it has different linguistic features than a domain that is
generated by a human. This assumption is reasonable because HGDs have
the primary purpose of being easily remembered and used by human
beings, thus are usually built in a way that meets this goal. On the
other hand, AGDs exhibit a certain degree of linguistic randomness, as
numerous samples of the same randomized algorithm exist. Corner cases
of this assumption are mitigated by the subsequent steps or otherwise
discussed in \S\ref{sec:discussion}.

\subsubsection{Linguistic Features}
\label{linguistic-features}
%\tod[All]{Queste feature linguistiche sono diverse da quelle calcolate
%  dagli altri. Qui sono funzione di un dominio, non di un GRUPPO di
%  domini. TODO: random groups vs. no split}

Given a domain $d$ and its prefix $p = p_d$, we extract two classes of
linguistic features to build a 4-element feature vector for each $d$.

\paragraph*{LF1: Meaningful Characters Ratio} This feature models the
ratio of characters of the chosen prefix $p$ that comprise a
meaningful word. Domains with a low ratio are likely automatically
generated.

Specifically, we split $p$ into $n$ meaningful subwords $w_i$ of at
least 3 symbols: $|w_i| \geq 3$, \hl{leaving out as few symbols as
  possible}:
\begin{equation*} R(d) = R(p) := \max \frac{\sum_{i=1}^n |w_i|}{|p|} \end{equation*}
In the case of $p = \mathtt{facebook}$, $R(p) = (|\mathtt{face}| +
|\mathtt{book}|)/8 = 1$, the prefix is fully composed of meaningful
words, whereas $p = \mathtt{pub03str}$, $R(p) = (|\mathtt{pub}|)/8 =
0.375$.

\paragraph*{LF2: $n$-gram Normality Score} This class of features
captures the pronounceability of the chosen prefix of a domain. This
problem is well studied in the field of linguistics, and can be
reduced to quantifying the extent to which a string adheres to the
phonotactics of the (English) language. The more permissible the
combinations of phonemes~\cite{scholes1966phonotactic,
  bailey2001determinants}, the more pronounceable a word is. Domains
with a scarce number of such combinations are likely automatically
generated.

We calculate this class of features by extracting the $n$-grams of
$p$, which are the substrings of $p$ of length $n \in \{1,2,3\}$, and
counting their occurrences in the (English) language dictionary. The
features are thus parametric to $n$:
\begin{equation*} S_n(d) = S_n(p) := \frac{\sum_{n\text{-gram }t\text{ in
}p}\mathrm{count}(t)}{|p|-n+1} \end{equation*}
where $\mathrm{count}\left(t\right)$ are the occurrences of the
$n$-gram $t$ in the dictionary.

Fig. \ref{fig:normality-score} shows the value of $S_2$, along with its
derivation, for one HGD and one AGD.
\begin{figure}[t]
  %bold tt
\centering
  \ttfamily\bfseries\scriptsize \begin{tabular}{ccccccc|c} fa & ac & ce & eb &
bo & oo & ok & \multirow{2}{*}{$S_2 = 170.8$}\\ $109$ & $343$ & $438$ & $29$ &
$118$ & $114$ & $45$ &  \\ \end{tabular} \vspace{0.2cm}
\begin{tabular}{ccccc|c} aa & aw & wr & rq & qv & \multirow{2}{*}{$S_2 =
13.2$}\\ $4$ & $45$ & $17$ & $0$ & $0$ &  \\ \end{tabular}

  %font back to normal
  \rmfamily\mdseries \caption{$2$-gram normality score $S_2$ for
\texttt{facebook.com} and \texttt{aawrqv.biz}.} \label{fig:normality-score}
\end{figure}

\subsubsection{Statistical Linguistic Filter} \thesystem uses
\textbf{LF1-2} to build a feature vector $\bm{f}(d) = [R(d),
S_{1,2,3}(d)]^T$. It extracts these features from a dataset of HGDs
(\textsf{\relsize{-.5}{Alexa}} top 100,000) and calculates their mean $\bm{\mu} =
\left[\overline{R}, \overline{S_1}, \overline{S_2},
  \overline{S_3}\right]^T$ and covariance (matrix) $\bm{C}$, \hl{which
  respectively represent the statistical average values of the
  features and their correlation. Strictly speaking, the mean defines
  the centroid of the HGD dataset in the features' space, whereas the
  covariance identifies the shape of the hyperellipsoid around the
  centroid containing all the samples. Our filter constructs a
  confidence interval, with the shape of such hyperellipsoid, that
  allows us to separate HGDs from AGDs with a measurable, statistical
  error that we set a priori.}

\hl{The rationale is that obtaining a dataset of HGDs is
  straightforward and does not constrain the filtering to specific
  AGDs: Our filter thus models non-AGDs by means of the generic
  modeling of HGDs.}

\paragraph*{Distance Measurement} To tell whether a previously unseen
domain $d'$ resembles the typical features of HGDs, the filter
measures the distance between the feature vector $\bm{f}(d') = \bm{x}$
and the centroid. To this end, we leverage the Mahalanobis distance:
$\dmah{\bm{x}} =
\sqrt{(\bm{x}-\bm{\mu})^T\bm{C}^{-1}(\bm{x}-\bm{\mu})}$. This distance
has the property of (1) taking into account the correlation between
features---which is significant, because of how the features are
defined, and (2) operating with scale-invariant datasets.

\paragraph*{Distance Threshold} A previously unseen domain $d'$ is
considered as automatically generated when its feature vector
identifies a point that is too distant from the centroid:
$\dmah{\bm{x}} > t$. To take a proper decision we define the threshold
$t$ as the $p$-percentile of the distribution of $\dmah{\bm{x}}$,
where $(1 - p)$ is the fraction of HGDs that we allow to confuse as
AGDs. In this way, we can set a priori the amount of errors.

As mentioned in \S\ref{agd-detection}, the \textbf{DGA Discovery}
module employs a strict threshold, $t = \Lambda$, whereas the
\textbf{AGD Detection} module requires a looser threshold, $t =
\lambda$, where $\lambda < \Lambda$.

\begin{figure}[t]
  \centering
  \includegraphics[width=.85\columnwidth]{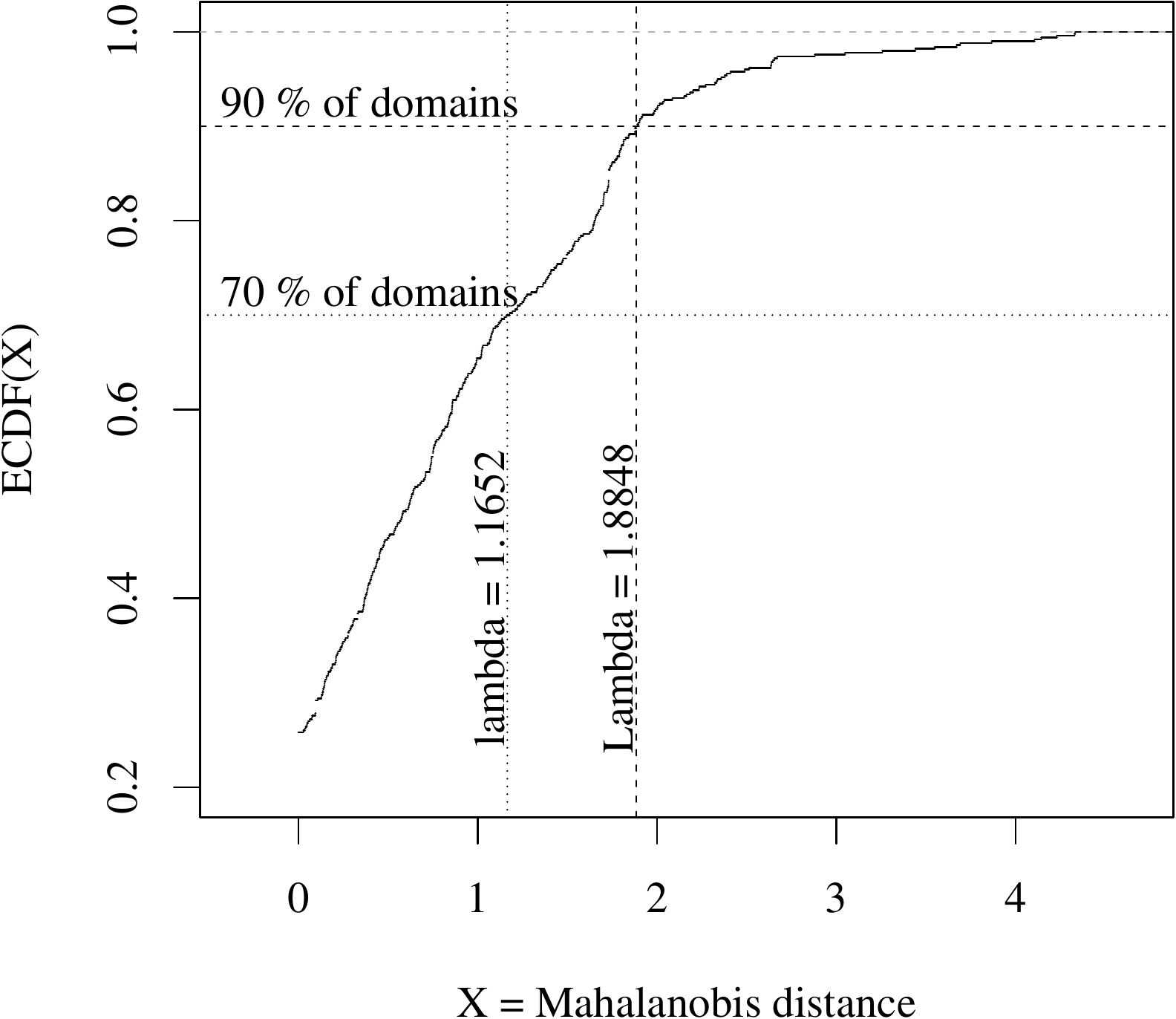}
  \caption{Mahalanobis distance ECDF for \textsf{\relsize{-.5}{Alexa}} top 100,000
    domains with $\lambda$ and $\Lambda$ identification.}
  \label{fig:ecdf-alexa}
\end{figure}

\begin{figure}[t]
  \centering
  \includegraphics[width=\columnwidth]{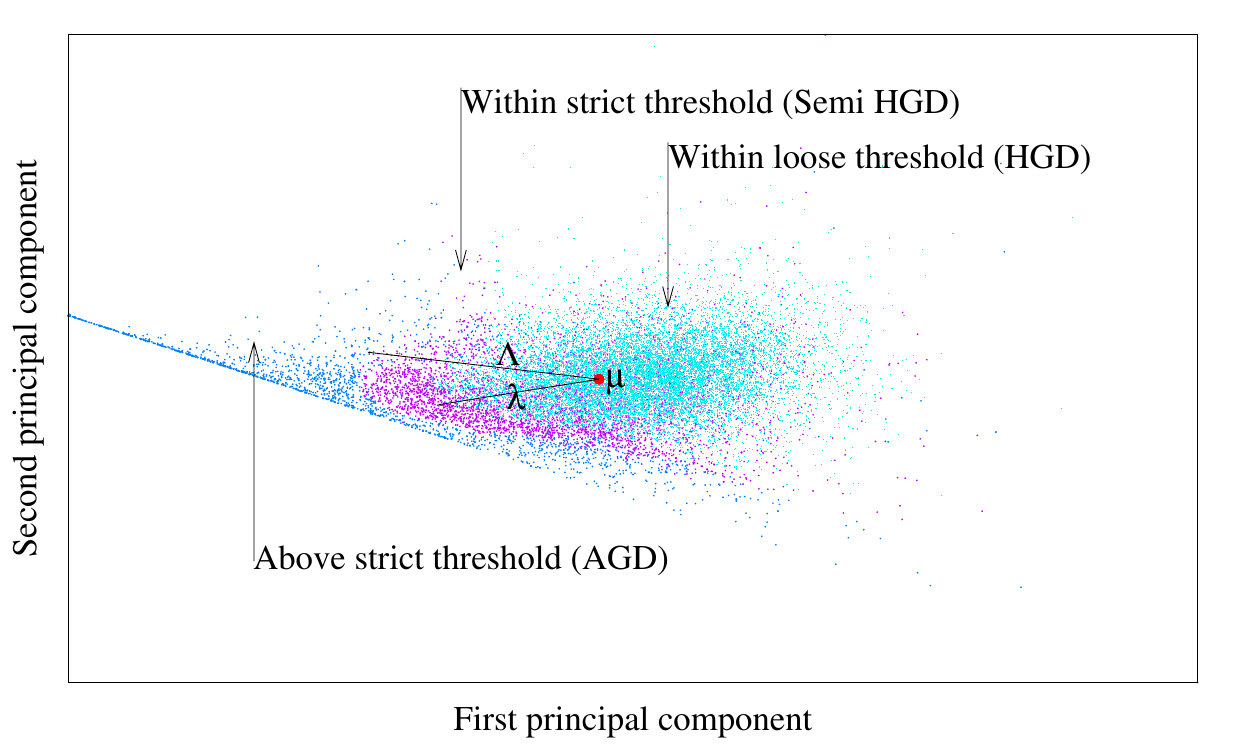}
  \caption{\hl{Principal components of the \textsf{\relsize{-.5}{Alexa}} top 100,000 domains hyperellipsoid with annotation of the confidence interval thresholds.}
  \label{fig:alexa-pca}}
\end{figure}

\paragraph*{Threshold Estimation}
\hl{To estimate proper values for $\lambda$ and $\Lambda$, we compute
  $\dmah{\bm{x}}$ for $\bm{x} = \bm{f}(d), \forall d \in
  \mathbb{D}_{HGD}$, whose distribution is plotted in}
Fig.~\ref{fig:ecdf-alexa}\hl{ as ECDF. We then set $\Lambda$ to the
  90-percentile and $\lambda$ to the 70-percentile of that
  distribution, as annotated in the figure.} Fig.~\ref{fig:alexa-pca}
\hl{depicts the 99\%-variance preserving 2D projection of the
  hyperellipsoid associated to $\mathbb{D}_{HGD}$, together with the
  confidence interval thresholds calculated as mentioned above.}

\subsection{Step~2: AGD Clustering} \label{sec:agd-clustering} This step
receives as input the set of domains $d \in \mathbb{D}$ that have passed
\textbf{Step~1}. These domains are such that $\dmah{\bm{f}(d)} > \Lambda$,
which means that $d$ is likely to be automatically generated, because it is too
far from the centroid of the HGDs.

The goal of this step is to cluster domains according to their
similarity. We define as \emph{similar} two domains that resolved to
similar sets of IP addresses. The rationale is that the botmaster of a
DGA-based botnet registers several domains that, at different points
in time, resolve to the same set of IPs (i.e., the C\&C servers). To
find similar domains, we represent the domain-to-IP relation as a
bipartite graph, which we convert in a proper data structure that
allows us to apply a spectral clustering
algorithm~\cite{Newman:2010wp} that returns the groups of similar
domains (i.e., nodes of the graph).

In this graph, two sets of node exists: $K=|\mathbb{D}|$ nodes represent the
domains, and $L = |\IPs(\mathbb{D})|$ nodes represent the IPs. An edge exists
from a node $d \in \mathbb{D}$ to node $l \in \IPs(\mathbb{D})$ whenever a
domain pointed to an IP.

\subsubsection{Bipartite Graph Recursive Clustering}
% \tod[All]{Throughout your discussion of features, it would be
% helpful to distinguish those features which are solely yours and
% those covered by previous work (e.g. Yadav, Antonakakis)}

\label{sec:bipartite-graph-recursive-clustering} To cluster the domain
nodes $\mathbb{D}$, we leverage the DBSCAN clustering
algorithm~\cite{han2006data}, which is fast and easy to implement in
our scenario.

\paragraph*{Data Structure} We encode the bipartite graph as a sparse
matrix $\bm{M} \in \mathbb{R}^{L \times K}$ with $L$ rows and $K$
columns. Each cell $M_{l,k}$ holds the weight of an edge $k \to l$ in
the bipartite graph, which represents the fact that domain $d_k$
resolves to IP $l$. The weight encodes the ``importance'' of this
relation. For each IP $l$ in the graph, the weights $M_{l,k}, \forall
k = 1, \dots, K$ are set to $\frac{1}{|\mathbb{D}(l)|}$, where
$\mathbb{D}(l) \subset \mathbb{D}$ is the subset of domains that point
to that IP. This weight encodes the peculiarity of each IP: The less
domains an IP is pointed by, the more characterizing it is.

\paragraph*{Domain Similarity} At this point we calculate the matrix
$\bm{S} \in \mathbb{R}^{K \times K}$, whose cells encode the
similarity between each pair of domains $d$ and $d'$. We want to
consider two domains as highly similar when they have peculiar IPs in
common. Therefore, we calculate the similarity matrix from the
weights, as $\bm{S} = \bm{N}^{T} \cdot \bm{N} \in \mathbb{R}^{K\times
  K}$, where $\bm{N}$ is basically $\bm{M}$ normalized by columns
(i.e., $\sum_{l=1}^L M_{l,k} = 1, \forall k = 1,K$). This similarity
matrix implements the rationale that we mentioned at the beginning of
this section.

\paragraph*{Domain Features and Clustering} We apply the DBSCAN
algorithm hierarchically. We compute the first normalized eigenvector
$\bm{v}$ from $\bm{S}$. At this point, each domain name $d_k$ can be
represented by its feature $v_k$, the $k$-th element of $\bm{v}$,
which is fed to the DBSCAN algorithm to produce the set of $R$
clusters $\mathcal{D} = \{\mathbb{D}^1, \dots, \mathbb{D}^R\}$ at the
current recursive step.

\paragraph*{Clustering Stop Criterion} We recursively repeat the
clustering process on the newly created clusters until one of the
following conditions is verified:

\begin{itemize}
\item a cluster of domains $\mathbb{D}' \in \mathcal{D}$ is \emph{too
    small} (e.g., it contains less than $25$ domains) thus it is
  excluded from the final result;
\item a cluster of domains has its $\bm{M}$ matrix with \emph{all the
    elements greater than zero}, meaning that the bipartite graph it
  represents is strongly connected;
\item a cluster of domains \emph{cannot be split further} by the
  DBSCAN algorithm with the value of $\epsilon$ set. In our
  experiments, we set $\epsilon$ to a conservative low value of $0.1$,
  so to avoid the generation of clusters that contain domains that are
  not similar.  Manually setting this value is possible because
  $\epsilon$, and the whole DBSCAN algorithm, works on normalized
  features.
\end{itemize}
The final output of DBSCAN is $\mathcal{D}^\star = \{\mathbb{D}^1,
\dots, \mathbb{D}^R\}$. The domains within each $\mathbb{D}^r$ are
similar among each other.

\subsubsection{Dimensionality Reduction}
\label{dimensionality-reduction}
The clustering algorithm employed has a space complexity of
$O(|\mathbb{D}|^2)$. To keep the problem feasible we randomly sample
our dataset $\mathbb{D}$ of AGDs into $I$ smaller datasets
$\mathbb{D}_i, i = 1, \dots, I$ of approximately the same size, and
cluster each of them independently, where $I$ is the minimum value
such that a space complexity in the order of $|\mathbb{D}_i|^2$ is
affordable. Once each $\mathbb{D}_i$ is clustered, we recombine the
$I$ clustered sets, $\mathcal{D}^\star_i = \{\mathbb{D}^1, \dots,
\mathbb{D}^{R_i}\}$, onto the original dataset $\mathbb{D}$. Note that
each $\mathbb{D}_i$ may yield a different number $R_i$ of
clusters. This procedure is very similar to the map-reduce programming
paradigm, where a large computation is parallelized into many
computations on smaller partitions of the original dataset, and the
final output is constructed when the intermediate results become
available.

We perform the recombination in the following post-processing phase,
which is run anyway, even if we do not need any dimensionality
reduction---that is, when $I = 1$ and thus $\mathbb{D}_1 \equiv
\mathbb{D}$.

\subsubsection{Clustering Post Processing}
\label{sec:post-processing}
We post process the set of clusters of domains $\mathcal{D}^\star_i,
\forall i$ with the following \textbf{Pruning} and \textbf{Merging}
procedures. For simplicity, we set the shorthand notation $\mathbb{A}
\in \mathcal{D}^\star_i$ and $\mathbb{B} \in \mathcal{D}^\star_j$ to
indicate any two sets of domains (i.e., clusters) that result from the
previous DBSCAN clustering, possibly with
$i=j$.%wo independent clusterings at reduced dimensionality on
%$\mathbb{D}_i, \mathbb{D}_j \subset \mathbb{D}$. Both $\mathbb{A}$ and
%$\mathbb{B}$ contain domains after the DBSCAN clustering.

\paragraph*{Pruning} Clusters of domains that exhibit a nearly
one-to-one relation with the respective IPs are considered unimportant
because, by definition, they do not reflect the concept of DGA-based
C\&Cs (i.e., many domains, few IPs). Thus, we filter out the clusters
that are flat and show a pattern-free connectivity in their bipartite
domain-IP representation. This allows to remove ``noise'' from the
dataset.

Formally, a cluster $\mathbb{A}$ is removed if $\frac{\left|
    \IPs\left(\mathbb{A}\right)\right|}{\left|\mathbb{A} \right|} >
\gamma$, where $\gamma$ is a threshold that is derived automatically
as discussed in \S\ref{sec:experiments}.

\paragraph*{Merging} Given two independent clusters $\mathbb{A}$ and
$\mathbb{B}$, they are merged together if the intersection between
their respective sets of IPs is not empty. Formally, $\mathbb{A}$ and
$\mathbb{B}$ are merged if $\IPs(\mathbb{A}) \cap \IPs(\mathbb{B})
\neq \emptyset$. This merging is repeated out iteratively, until every
combination of two clusters violates the above condition.

\medskip\noindent The outcome of the post-processing phase is thus a
set of clusters of domains $\mathcal{E} = \{\mathbb{E}^1, \dots,
\mathbb{E}^Q\}$ where each $\mathbb{E}^q$ (1) exhibits a domain-to-IP
pattern and (2) is disjunct to any other $\mathbb{E}^p$ with respect
to its IPs. In conclusion, each cluster $\mathbb{E}$ contains the AGDs
employed by the same botnet backed by the C\&C servers at IP addresses
$\IPs(\mathbb{E})$.

\subsection{Step~3: DGA Fingerprinting}
\label{sec:dga-fingerprinting-and-new-agd-detection} The AGD clusters
identified with the previous processing are used to extract
fingerprints of the DGAs that generated them. In other words, the goal
of this step is to extract the invariants of a DGA. We use these
fingerprints in the \textbf{AGD Detection} module to assign labels to
previously unseen domains, if they belong to one of the clusters.

Given a generic AGD cluster $\mathbb{E}$, corresponding to a given
DGA, we extract the following cluster features:

\paragraph*{CF1: C\&C Servers Addresses} Defined as $\IPs(\mathbb{E})$.

\paragraph*{CF2: Chosen Prefix Length Range} Captures the lengths of
the chosen prefix allowed for the domains in $\mathbb{E}$. The
boundaries are defined as the lengths of the shortest and longest
chosen prefixes of the domains of $\mathbb{E}$.

\paragraph*{CF3: Chosen Prefix Character Set} $C$ employed for the
chosen prefixes of the domains, defined as $C := \bigcup_{e \in
  \mathbb{E}} \mathrm{charset}(p_e)$, where $p_e$ is the chosen prefix
of $e$. It captures what characters are used during the random
generation of the domain names.

\paragraph*{CF4: Chosen Prefix Numerical Characters Ratio Range}
$[r_m, r_M]$ captures the ratio of numerical characters allowed in the
chosen prefix of a given domain. The boundaries are, respectively, the
minimum and the maximum of $\frac{\mathrm{num}(p_e)}{|p_e|}$ within
$\mathbb{E}$, where $\mathrm{num}(p_e)$ is the number of numerical
characters in the chosen prefix of $e$.

\paragraph*{CF5: Public Suffix Set} The set of eTDL employed by the domains in
$\mathbb{E}$.

\medskip\noindent To some extent, these features define the
aposteriori linguistic characteristics of the domains found within
each cluster $\mathbb{E}$. In other words, they define a model of
$\mathbb{E}$.

\subsection{AGD Detection Module}
\label{sec:agd-detection}
This module receives a previously unseen domain $d$ and decides
whether it is automatically generated by running the \textbf{AGD
  Filtering} step with a loose threshold $\lambda$. If $d$ is
automatically generated, it is matched against the fingerprints of the
known DGAs on the quest for correspondences.

In particular, we first select the candidate AGD clusters
$\{\mathbb{E}\}$ that have at least one IP address in common with the
IP addresses that $d$ pointed to: $\IPs(d) \cap \IPs(\mathbb{E}) \neq
\emptyset, \forall \mathbb{E}$. Then, we select a subset of candidate
clusters such that have the same features \textbf{CF1--5} of
$d$. Specifically, the length of the chosen prefix of $d$, its
character set, its numerical characters ratio, and the eTLD of $d$
must lie within the ranges defined above.

The clusters that survive this selection are chosen as the labels of $d$.

\begin{figure*}[t]
  \centering
  \includegraphics[width=.7\textwidth]{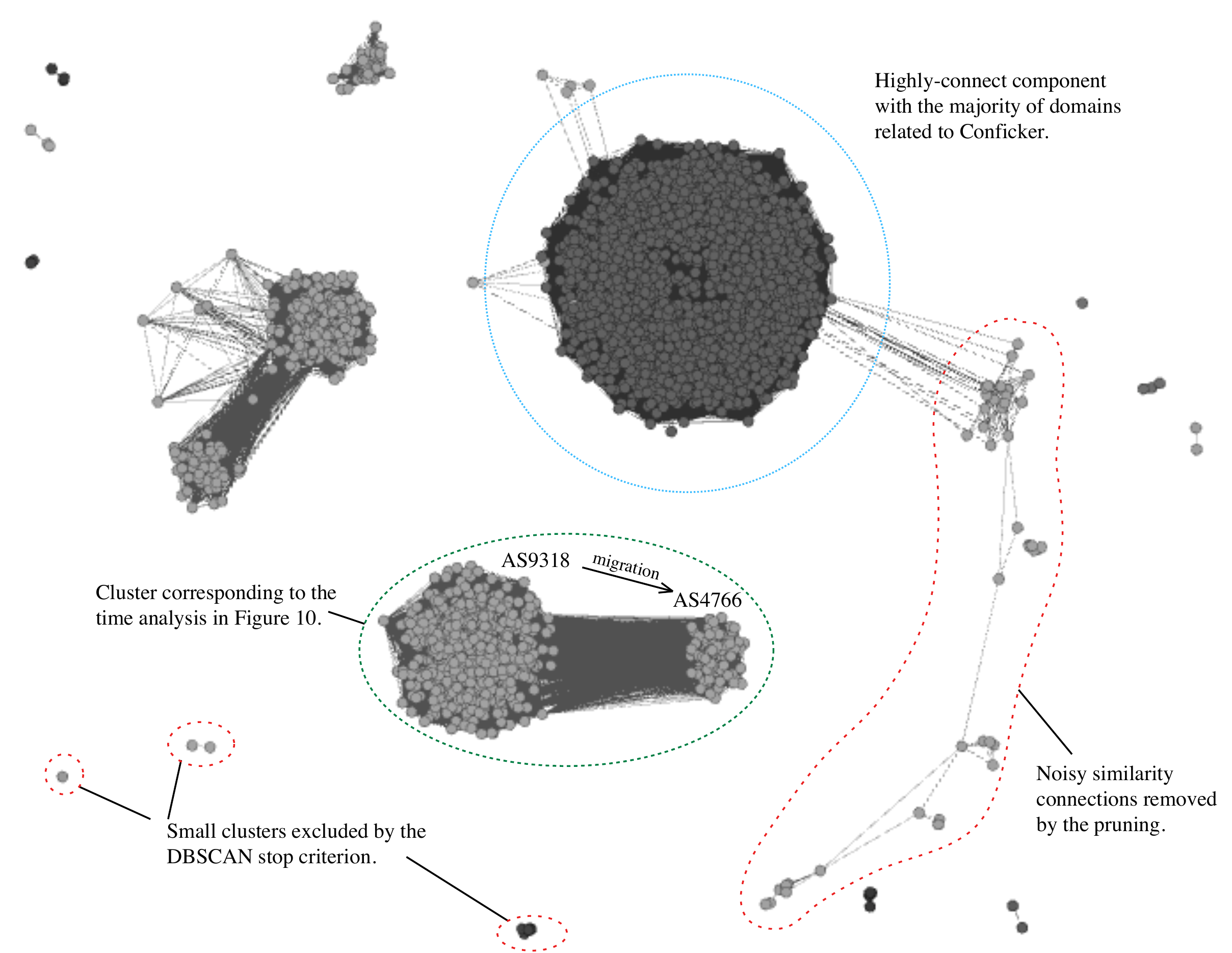}
  \caption{Graph representation of the similarity matrix $\bm{S}$
    during the first run of the DBSCAN clustering algorithm, as
    defined in \S\ref{sec:bipartite-graph-recursive-clustering}. The
    nodes represent the domains, while edges represent the similarity
    relations. The purpose of the DBSCAN is to isolate
    strongly connected ``communities'' of domains from ``noisy'' and
    uncorrelated domains.}
  \label{fig:clusters}
\end{figure*}

\section{Experimental Evaluation}
\label{sec:experiments}
Validating the results of the \thesystem is not trivial, because it
produces novel knowledge. The reason is that we do not have a ground
truth available that would allow us to strictly validate the
correctness of \thesystem quantitatively. For example, no information
is available, to our knowledge, about the membership of a malicious
domain to one family of AGDs. If such ground truth were available,
then there would be no need for \thesystem.

In lack of an established ground truth, we proceed as follows. We
validate \emph{quantitatively} the internal components of each module
(e.g., to verify that they do not produce meaningless results and to
assess the sensitivity of the parameters), and \emph{qualitatively}
the whole approach, to make sure that it produces useful knowledge
with respect to publicly available information.

\subsection{Evaluation Dataset and Setup}
\label{sec:datas-descr-exper}
The \textbf{DGA Discovery} module of \thesystem requires a feed of
\hl{recursive} DNS traffic and a reputation system that tells whether
a domain is generally considered as malicious. For the former data
source, we obtained access to the ISC/SIE framework~\cite{isc_sie},
which provides DNS traffic data shared by hundreds of different
network operators. Differently from~\cite{antonakakisthrow}, this type
of traffic is privacy preserving and very easy to collect. For the
latter data source we used the
\textsf{\relsize{-.5}{Exposure}}~\cite{bilge2011exposure} blacklist,
which included $107,179$ distinct domains as of October 1st, 2012.

Differently from~\cite{yadav2010detecting}, we used AGDs merely as a
ground truth for validation, not for bootstrapping our system before
run time. More precisely, to \emph{validate} the components of
\thesystem we \hl{relied on ground truth generated by} publicly
available implementations of the DGAs used by
Conficker~\cite{leder2009know} and
Torpig~\cite{Stone-Gross:2009:YBM:1653662.1653738}, which have been
among the earliest and most widespread botnets that relied on DGAs for
C\&C communication. After Conficker and Torpig, the use of DGAs kept
rising. With these DGAs we generated five datasets of domains, which
resemble (and in some cases are equivalent to) the domains generated
by the actual botnets: 7500, 7750 and 1,101,500 distinct AGDs for the
\textbf{Conficker.A}, \textbf{Conficker.B} and \textbf{Conficker.C}
malware, respectively, and 420 distinct AGDs for the \textbf{Torpig}
dataset. Moreover, we collected the list of 36,346 AGDs that
\textsf{\relsize{-.5}{Microsoft}} claimed in early 2013 to be related to the activity
of \textbf{Bamital}\footnote{\url{http://noticeofpleadings.com/}}.

% To our knowledge, no extensive dataset is available to associate to these
% AGDs their relative DNS traffic information.

We ran our experiments on a 4-core machine equipped with 24GB of
physical memory. All the runs required execution times in the order of
the minutes.

\subsection{DGA Discovery Validation}
\label{sec:dga-discovery-validation}

\subsubsection{Step~1: AGD Filtering} The AGD filter is used in two
contexts: by the \textbf{DGA Discovery} module as a pre-clustering
selection to recognize the domains that appear automatically generated
within a feed of malicious domains, and by the \textbf{AGD Detection}
module as a pre-labeling selection. For pre-clustering, the strict
threshold $\Lambda$ is enforced to make sure that no non-AGD domains
pass the filter and possibly bias the clustering, whereas for
pre-labeling the loose threshold $\lambda$ is used to allow more
domains to be labeled. Recall that, the \textbf{Labeler} will
eventually filter out the domains that resemble no known AGD. We test
this component in both the contexts against the AGD datasets of
\textbf{Conficker}, \textbf{Torpig} and \textbf{Bamital} (that we had
never seen before).

%\tod[All]{In the same vein, how does your system cope with DGAs
%  adapting and using more english-like domains? Its worth discussing
%  (and even better experimentally validating) how susceptible your
%  approach is to evasion. -- Non ne esistono!}

\begin{figure}[t] \centering
\includegraphics[width=.9\columnwidth]{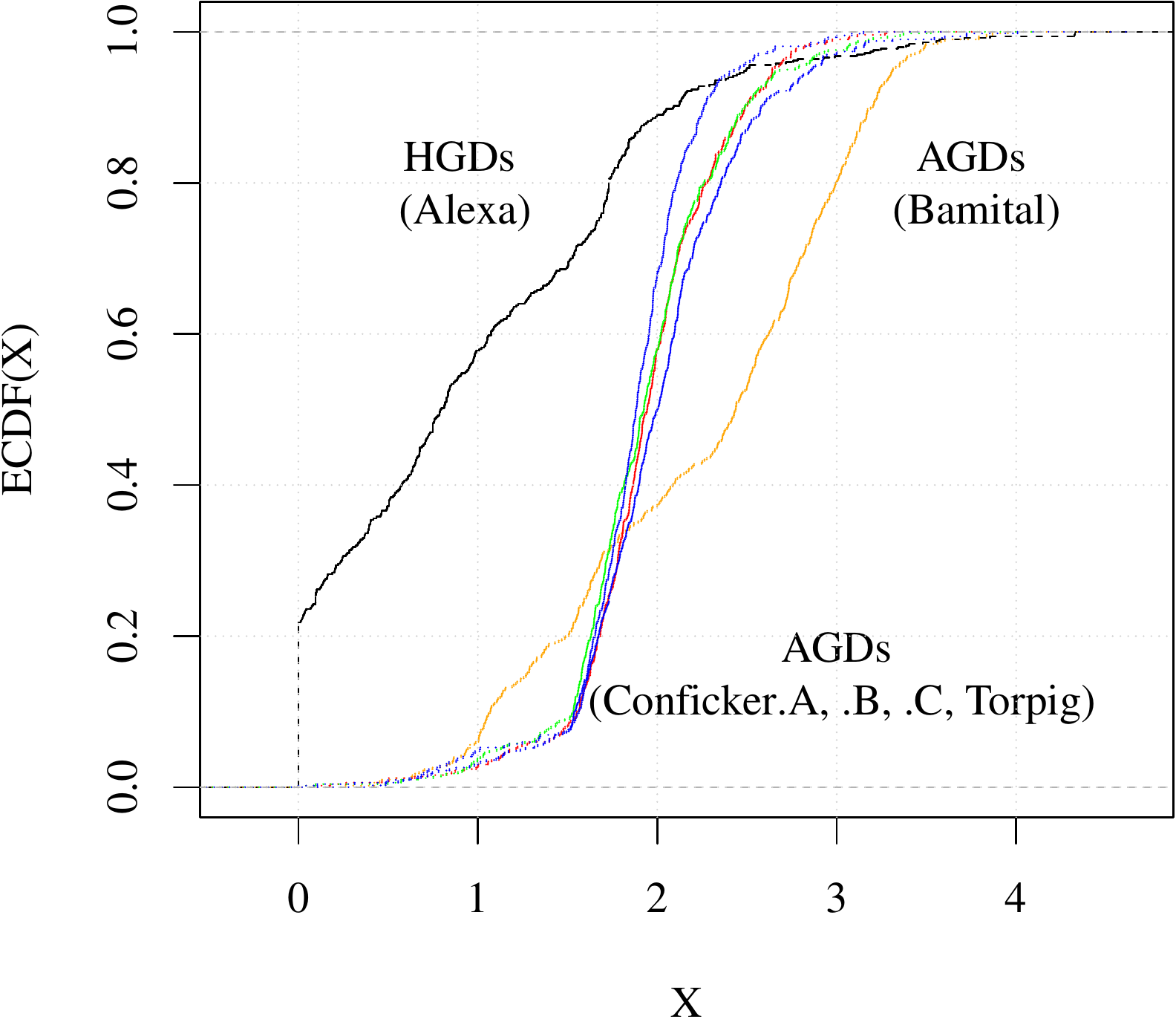}
\caption{Mahalanobis distance ECDF for different datasets. A KS
  statistical test to compare distinct AGD distributions against the
  HGD distribution yields $p$-values close to zero, confirming that
  they are drawn from diverse
  processes.} \label{fig:ecdf-agd} \end{figure}

The filter, which is the same in both the contexts, is best visualized
by means of the ECDF of the Mahalanobis
distance. Fig.~\ref{fig:ecdf-agd} shows the ECDF from the AGD
datasets, compared to the ECDF from the \textsf{\relsize{-.5}{Alexa}} top 100,000
domains. The plot shows that each datasets of AGDs and HGDs have
different distribution: This confirms that our linguistic features are
well suited to perform the discrimination. Interestingly, each AGD
dataset has a distinctive distribution: \textbf{Conficker} and
\textbf{Torpig} have the same linguistic features, which differ from
the linguistic features of \textbf{Bamital} (and, of course, from the
non-AGD domains). We may argue that the DGAs that are responsible of
such characteristic features are also different, whereas
\textbf{Conficker} and \textbf{Torpig} rely on DGAs with very similar
linguistic characteristics.
Then, we verify which fraction of AGDs passes the filter and reaches
the \textbf{AGD Clustering} ($\Lambda$) step or the \textbf{Labeler}
($\lambda$). The results obtained are reported in the first column of
Tab.~\ref{tab:agd-filtering} and show that roughly half of the domains
would not contribute to the generation of the clusters: The
conservative settings ensure that only the domains that exhibit the
linguistic features more remarkably are used for
clustering. Ultimately, most of the true AGD domains will be labeled
as such before reaching the \textbf{Labeler}. Overall, \thesystem has
a recall of 81.4 to 94.8\%, which is remarkable for a non-supervised
and completely automatic approach.
\begin{table}[b]
  \centering
  \begin{tabular}{rc>{\bfseries}c}
    \toprule
    \multirow{2}{*}{\textsc{Malware}} & $\bm{d_{Mah} > \Lambda}$  & $\bm{d_{Mah} >
      \lambda}$\\

    \cmidrule{2-3}

    & Pre-clustering selection & {\mdseries Recall}\\

    \midrule

    Conficker.A & 46.5\% & 93.4\% \\

    Conficker.B & 47.2 \% & 93.7\% \\

    Conficker.C & 52.9 \% & 94.8\% \\

    Torpig & 34.2\% & 93.0\% \\

    Bamital & 62.3\% & 81.4\% \\

    \bottomrule
  \end{tabular}
  \caption{AGD pre-clustering selection and
    recall.}
  \label{tab:agd-filtering} \end{table}
%
%\vspace{-1.5em}
\subsubsection{Step~2: AGD Clustering}
\label{sec:eval-clustering}
We ran \thesystem on our dataset and, after the first run of the
DBSCAN clustering, we obtained the similarity matrix depicted in
Fig.~\ref{fig:clusters}. Even with one run of the algorithm, we can
already see some interesting groups of domains that are similar. The
annotations on the figure are clarified in the reminder of this
section.

\paragraph*{Reality Check} We searched for qualitative ground truth
that could confirm the usefulness of the clusters obtained by running
\thesystem on our dataset. To this end, we queried \textsf{\relsize{-.5}{Google}} for
the IP addresses of each AGD cluster to perform manual labeling of
such clusters with evidence about the malware activity found by other
researchers.

We gathered evidence about a cluster with $33,771$ domains allegedly
used by \textsf{\relsize{-.5}{Conficker}} (see also Fig.~\ref{fig:clusters}) and
another cluster with $3870$ domains used by \textsf{\relsize{-.5}{Bamital}}. A
smaller cluster of $392$ domains was assigned to \textsf{\relsize{-.5}{SpyEye}}, and
two clusters of $404$ and $58$ domains, respectively, were assigned to
\textsf{\relsize{-.5}{Palevo}}. We were unable to find information to label the
remaining six clusters as related to known malware.

This reality check helped us confirming that we successfully isolated
domains related to botnet activities and IP addresses hosting C\&C
servers. The remainder of this section evaluates how well such
isolation performs in general settings (i.e., not on a specific
dataset).

\begin{figure}
  \centering
  \includegraphics[width=\columnwidth]{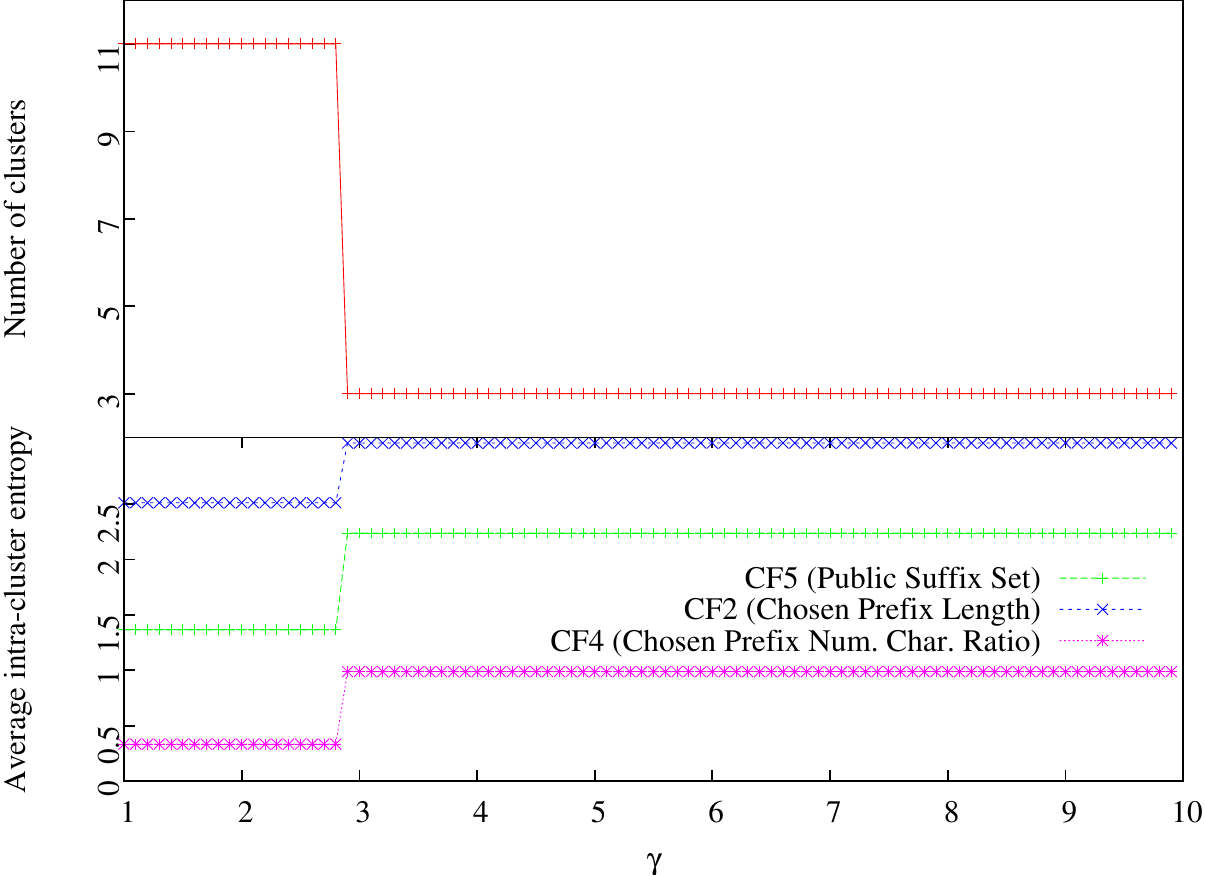}
  \caption{Clustering sensitivity from parameter $\gamma$. By studying
    the number of clusters (top) and the average intra-cluster entropy
    over \textbf{CF2}, \textbf{4}, \textbf{5} (bottom), we can choose
    the best $\gamma \in (0, 2.8)$.}
  \label{fig:gamma-sensitivity}
\end{figure}

\paragraph*{Sensitivity From $\bm{\gamma}$} We evaluated the
sensitivity of the clustering result to the $\gamma$ threshold used
for cluster pruning. To this end, we studied the number of clusters
generated with varying values of $\gamma$. A steady number of cluster
indicates low sensitivity from this parameter, which is a desirable
property. Moreover, abrupt changes of the number of clusters caused by
certain values of $\gamma$ can be used as a decision boundary to this
parameter. Fig.~\ref{fig:gamma-sensitivity} shows such a change at
$\gamma = 2.8$.

We also assessed how $\gamma$ influences the quality of the clustering
to find safety bounds of this parameter within which the resulting
clusters do not contain spurious elements. In other words, we want to
study the influence of $\gamma$ on the cluster features calculated
within each cluster. To this end, we consider the cluster features for
which a simple metric can be easily defined: \textbf{CF2 (Chosen
  Prefix Length Range)}, \textbf{CF4 (Chosen Prefix Numerical
  Characters Ratio Range)} and \textbf{CF5 (Public Suffix Set)}. A
clustering quality is high if all the clusters contain domains that
are uniform with respect to these features (e.g., each cluster contain
elements with common public suffix set or length). We quantify such
``uniformity'' as the entropy of each features. As
Fig.~\ref{fig:gamma-sensitivity} shows, all the features reflect an
abrupt change in the uniformity of the clusters around $\gamma = 2.8$,
which corroborates the above finding.

In conclusion, values of $\gamma$ outside $(0,2.8)$ do not allow the clustering
algorithm to optimally separate clusters of domains.

\paragraph*{Correctness} Our claim is that the clustering can
distinguish between domains generated by different DGAs by means of
the representative IPs used by such DGAs (which are likely to be the
C\&C servers). To confirm this claim in a robust way, we evaluate the
quality of the clustering with respect to features other than the IP
addresses. In this way, we can show that our clustering tells
different DGAs apart, regardless of the IP addresses in common. In
other words, we show that our clustering is independent from the
actual IP addresses used by the botnets but it is capable of
recognizing DGAs in general.

To this end, we ignore \textbf{CF1} and calculate the features
\textbf{CF2-5} of each cluster and show that they are distributed
differently between any two clusters. We quantify this difference by
means of the $p$-value of the Kolmogorov-Smirnov (KS) statistical
test, which tells how much two samples (i.e., our \textbf{CF2-5}
calculated for each couple of clusters) are drawn from two different
stochastic processes (i.e., they belong to two different
clusters). $p$-values toward 1 indicate that two clusters are not well
separated, because they comprise domains that are likely drawn from
the same distribution. On the other hand, $p$-values close to zero
indicate sharp separation.

The results summarized in Tab.~\ref{tab:pvalues} confirm that most of
the clusters are well separated, because their $p$-value is close to
0. In particular 9 of our 11 clusters are highly dissimilar, whereas
two clusters are not distinguishable from each other (Clusters 2 and
4). From a manual analysis of these two clusters we can argue that a
common DGA is behind both of them, even if there is no strong evidence
(i.e. DNS features) of this being the case.  Cluster 2 include domains
such as \texttt{46096.com} and \texttt{04309.com}, whereas two samples
from Cluster 4 are \texttt{88819.com} and \texttt{19527.com}.

%
% PERICOLOSO: stiamo di fato dicendo che con meno feature separiamo meglio
% potrebbero chiedersi come mai e, soprattutto, potrebbero dire che speculiamo
% su una feature troppo semplice come l'eTLD
%
% We confirm the previous results by replicating the experiments employing the
% eTDL of the domains in the clusters. For each pair of clusters $E_i$ and
% $E_j$ we compute the mean entropy of the clusters using the eTLD as class
% labels. Then, we compare the result with the entropy of $E_i \cup E_j$. The
% results show non-negative information gains, meaning that the splits are
% properly done.
%

\begin{table}[b]
  \centering
  \scriptsize
  \setlength{\tabcolsep}{.3em}

  \begin{tabular}{c c c c c c c c c c c} \cline{1-2} \multicolumn{1}{|c|}{2} &
    \multicolumn{1}{c|}{e-12}\\

 \cline{1-3} \multicolumn{1}{|c|}{3} &
    \multicolumn{1}{c|}{e-8} & \multicolumn{1}{c|}{e-9}\\

 \cline{1-4}
    \multicolumn{1}{|c|}{4} & \multicolumn{1}{c|}{e-12} & \multicolumn{1}{c|}{1.00}
    & \multicolumn{1}{c|}{e-9}\\

 \cline{1-5} \multicolumn{1}{|c|}{5} &
    \multicolumn{1}{c|}{e-26} & \multicolumn{1}{c|}{e-32} &
    \multicolumn{1}{c|}{e-5} & \multicolumn{1}{c|}{e-36}\\

 \cline{1-6}
    \multicolumn{1}{|c|}{6} & \multicolumn{1}{c|}{e-33} & \multicolumn{1}{c|}{e-39}
    & \multicolumn{1}{c|}{e-27} & \multicolumn{1}{c|}{e-44} &
    \multicolumn{1}{c|}{0.00}\\

 \cline{1-7} \multicolumn{1}{|c|}{7} &
    \multicolumn{1}{c|}{e-3} & \multicolumn{1}{c|}{e-4}  &
    \multicolumn{1}{c|}{0.01} & \multicolumn{1}{c|}{e-3}  &
    \multicolumn{1}{c|}{e-16} & \multicolumn{1}{c|}{e-33}\\

 \cline{1-8}
    \multicolumn{1}{|c|}{8} & \multicolumn{1}{c|}{e-20} & \multicolumn{1}{c|}{e-11}
    & \multicolumn{1}{c|}{0.14} & \multicolumn{1}{c|}{e-12} &
    \multicolumn{1}{c|}{e-55} & \multicolumn{1}{c|}{e-276} &
    \multicolumn{1}{c|}{e-5}\\

 \cline{1-9} \multicolumn{1}{|c|}{9} &
    \multicolumn{1}{c|}{e-18} & \multicolumn{1}{c|}{e-5} & \multicolumn{1}{c|}{e-4}
    & \multicolumn{1}{c|}{e-5} & \multicolumn{1}{c|}{e-142} &
    \multicolumn{1}{c|}{e-305} & \multicolumn{1}{c|}{e-4} &
    \multicolumn{1}{c|}{e-16}\\

 \cline{1-10} \multicolumn{1}{|c|}{10} &
    \multicolumn{1}{c|}{e-14} & \multicolumn{1}{c|}{e-23} &
    \multicolumn{1}{c|}{e-19} & \multicolumn{1}{c|}{e-24} &
    \multicolumn{1}{c|}{e-50} & \multicolumn{1}{c|}{e-52} &
    \multicolumn{1}{c|}{e-17} & \multicolumn{1}{c|}{e-46} &
    \multicolumn{1}{c|}{e-46}\\

 \cline{1-11} \multicolumn{1}{|c|}{11} &
    \multicolumn{1}{c|}{e-22} & \multicolumn{1}{c|}{e-28} &
    \multicolumn{1}{c|}{0.61} & \multicolumn{1}{c|}{e-31} &
    \multicolumn{1}{c|}{e-163} & \multicolumn{1}{c|}{0.00} &
    \multicolumn{1}{c|}{e-8} & \multicolumn{1}{c|}{e-27} &
    \multicolumn{1}{c|}{e-82} & \multicolumn{1}{c|}{e-52}\\

 \cline{1-11} &
    \multicolumn{1}{|c|}{1} & \multicolumn{1}{c|}{2} & \multicolumn{1}{c|}{3} &
    \multicolumn{1}{c|}{4} & \multicolumn{1}{c|}{5} & \multicolumn{1}{c|}{6} &
    \multicolumn{1}{c|}{7} & \multicolumn{1}{c|}{8} & \multicolumn{1}{c|}{9} &
    \multicolumn{1}{c|}{10}\\

 \cline{2-11}
  \end{tabular}
\caption{The low
  $p$-values of the pairwise KS test between the lengths of the chosen prefix of
  the elements within each couple of clusters indicate that the elements of each
  pair of clusters have diverse features (we hereby exemplify the length of the
  chosen prefixes for the purpose of visualization). Thus, the clusters are well
  separated.}
\label{tab:pvalues}
\end{table}

\subsection{AGD Detection Evaluation}
We want to evaluate qualitatively how well the \textbf{AGD Detection}
module is able to assign the correct labels to previously unseen
suspicious domains. To this end, we first run the \textbf{AGD
  Discovery} module using the historical domain-to-IP relations
extracted from the ISC/SIE database for the domains indicated as
generically malicious by the malicious domain filter (which is
\textsf{\relsize{-.5}{Exposure}} in our case). Once this module produced the
clusters, we validated the outcome of the \textbf{AGD Detection}
against another (random) split of the same type of data extracted from
the ISC/SIE dataset (never observed before).

The result of the \textbf{AGD Detection} is a list of previously
unseen domains, assigned to a cluster (i.e., a DGA). Some examples of
previously unseen domains are depicted in Fig.~\ref{fig:previously
  unseen} along with some samples of the clusters where they have been
assigned to.

These examples show that \thesystem is capable of assigning the
correct cluster to unknown suspicious domains. Indeed, despite the
variability of the eTLD, which is commonly used as anecdotal evidence
to discriminate two botnets, our system correctly models the
linguistic features and the domain-to-IP historical relations and
performs a better labeling. In the second case the domains were all
registered under \texttt{.cn}, and it is also clear that they share
the same generation mechanism.
\begin{figure}
  \centering
  \scriptsize

\begin{tabular}{>{\ttfamily}c>{\ttfamily}c>{\ttfamily}c>{\ttfamily}c}
\multicolumn{4}{c}{Previously unseen domains} \\
 \hline hy613.cn & 5ybdiv.cn &
73it.cn & 39yq.cn \\
 69wan.cn & hy093.cn & 08hhwl.cn & hy267.cn \\
 hy673.cn &
onkx.cn & xmsyt.cn & fyf123.cn \\
 watdj.cn & dhjy6.cn & algxy.cn & g3pp.cn \\
\end{tabular}
\vspace{0.2cm} \\

\begin{tabular}{|>{\ttfamily}c>{\ttfamily}c>{\ttfamily}c>{\ttfamily}c|}
\multicolumn{4}{c}{Cluster 9 (Palevo)} \\
 \hline pjrn3.cn & 3dcyp.cn & x0v7r.cn
& 0iwzc.cn\\
 0bc3p.cn & hdnx0.cn & 9q0kv.cn & 4qy39.cn\\
 5vm53.cn & 7ydzr.cn &
fyj25.cn & m5qwz.cn\\
 qwr7.cn & xq4ac.cn & ygb55.cn & v5pgb.cn\\
 \hline
\end{tabular}
\vspace{0.3cm} \\

\begin{tabular}{>{\ttfamily}c>{\ttfamily}c>{\ttfamily}c>{\ttfamily}c}
\multicolumn{4}{c}{Previously unseen domains} \\
 \hline dky.com & ejm.com &
eko.com & blv.com\\
 efu.com & elq.com & bqs.com & dqu.com\\
 bec.com & dpl.com &
eqy.com & dyh.com\\
 dur.com & bnq.com & ccz.com & ekv.com\\
 \end{tabular}
\vspace{0.2cm} \\

\begin{tabular}{|>{\ttfamily}c>{\ttfamily}c>{\ttfamily}c>{\ttfamily}c|}
\multicolumn{4}{c}{Cluster 10 (Palevo)} \\
 \hline uon.org & jhg.org & eks.org &
kxc.com\\
 mzo.net & zuh.com & bwn.org & khz.net\\
 zuw.org & ldt.org & lxx.net &
epu.org\\
 ntz.com & cbv.org & iqd.com & nrl.net\\
 \hline \end{tabular}

  \caption{Labeling of previously unseen domains (see Appendix A).}
\label{fig:previously unseen} \end{figure}

\subsection{Intelligence and Insights}
\label{sec:intellicence-eval}
In this section, we describe two use cases of the \textbf{Intelligence
  and Insights} module, which provides the analyst with valuable
knowledge from the outputs of the other modules. The correctness of
the conclusions drawn from this module is predicated on the
correctness of the two upstream modules, already discussed in prevoius
sections.

\begin{figure}[t]
  \includegraphics[width=\columnwidth]{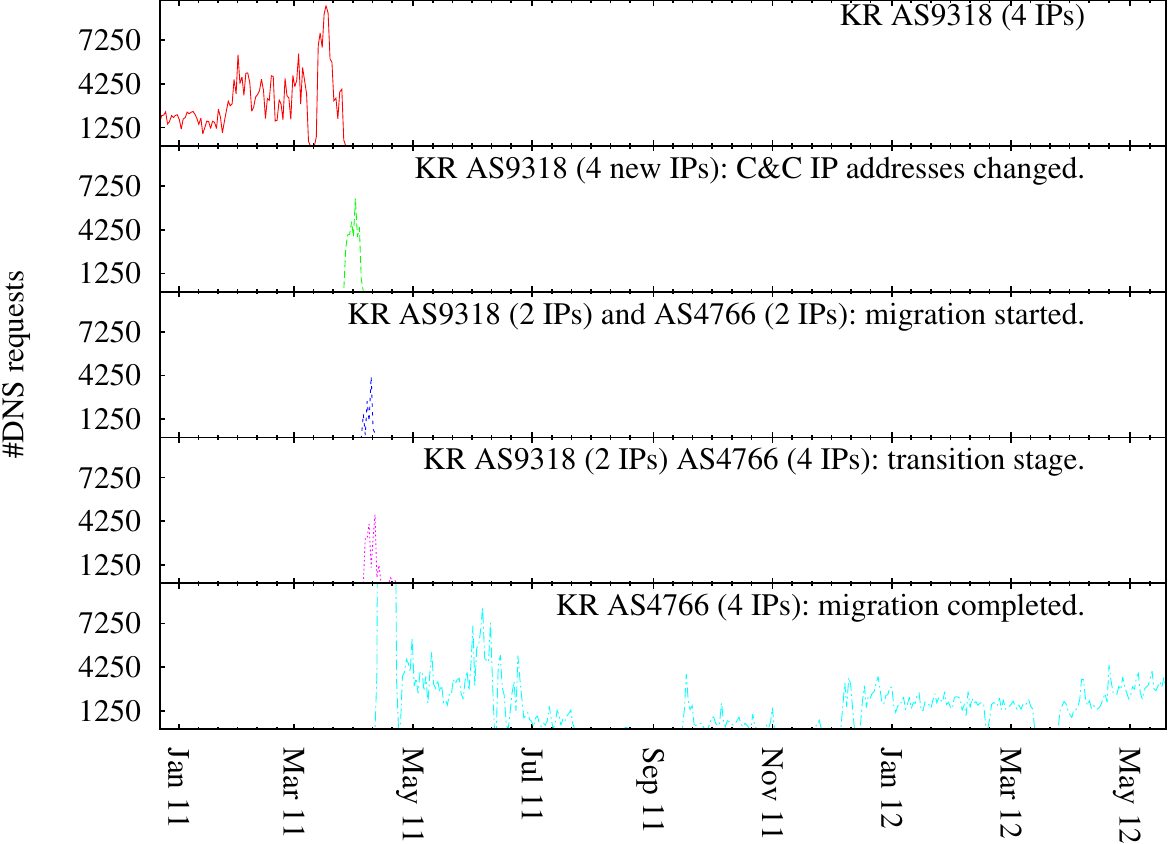}
  \caption{\textbf{Bamital}: Migration of C\&C from AS9318 to AS4766.}
  \label{fig:time-evolution-migration}
\end{figure}

\paragraph*{Unknown DGA Recognition From Scarce Data} Our system is
designed to automatically label the malicious AGDs related to
botnet activities. This is done by using the information of the DNS
traffic related to them.  Interestingly, some conclusions can be drawn
on previously unseen domains even in the unlucky case that such
information is missing (i.e., when no DNS data is available).

While asking on mailing lists for information about sinkholed IPs, we
received an inquiry by a group of researchers on February 9th. They
had found a previously unseen list of AGDs which resembled no known
botnet. Such list was the only information that they provided us
with. \thesystem labeled these domains with the fingerprints of a
Conficker cluster. This allowed the researchers to conduct further
investigation, which eventually confirmed that it was Conficker.B.

In conclusion, starting from the sole knowledge of a list of malicious
domains that \thesystem had never seen before, we discovered that,
according to our datasets, the only DGA able to produce domains with
that linguistic features was the DGA associated with
\textbf{Conficker}.

\paragraph*{Time Evolution} Associating AGDs to the activity of a
specific botnet allows to gather further information on that botnet,
by using the DGA fingerprints as a ``lookup index'' to make precise
queries.  We can track the behavior of a botnet to study its evolution
over time.

For instance, given a DGA fingerprint or AGD sample, we can select the
domains of the corresponding cluster $\mathbb{E}_{DGA}$ and partition
this set at different granularity (e.g., IPs or ASs) by considering
the \textsl{exact} set of IPs (or ASs) that they point to. Given the
activity that we want to monitor, for instance, the DNS traffic of
that botnet, we can then plot one time series for each partition.  In
our example, we count the number of DNS requests seen for the domains
in that partition at a certain sampling frequency (e.g., daily). The
analysis of the stacked time series generated allows to draw
conclusion about the behavior over time of the botnet.
Fig.~\ref{fig:time-evolution-migration} shows the case of (a) a
migration (the botmaster moved the C\&C servers from one AS to
another) followed by (b) a load balancing change in the final step
(the botmaster shut down 2 C\&C servers thus reducing the load
balancing).

\begin{figure}[t]
  \includegraphics[width=\columnwidth]{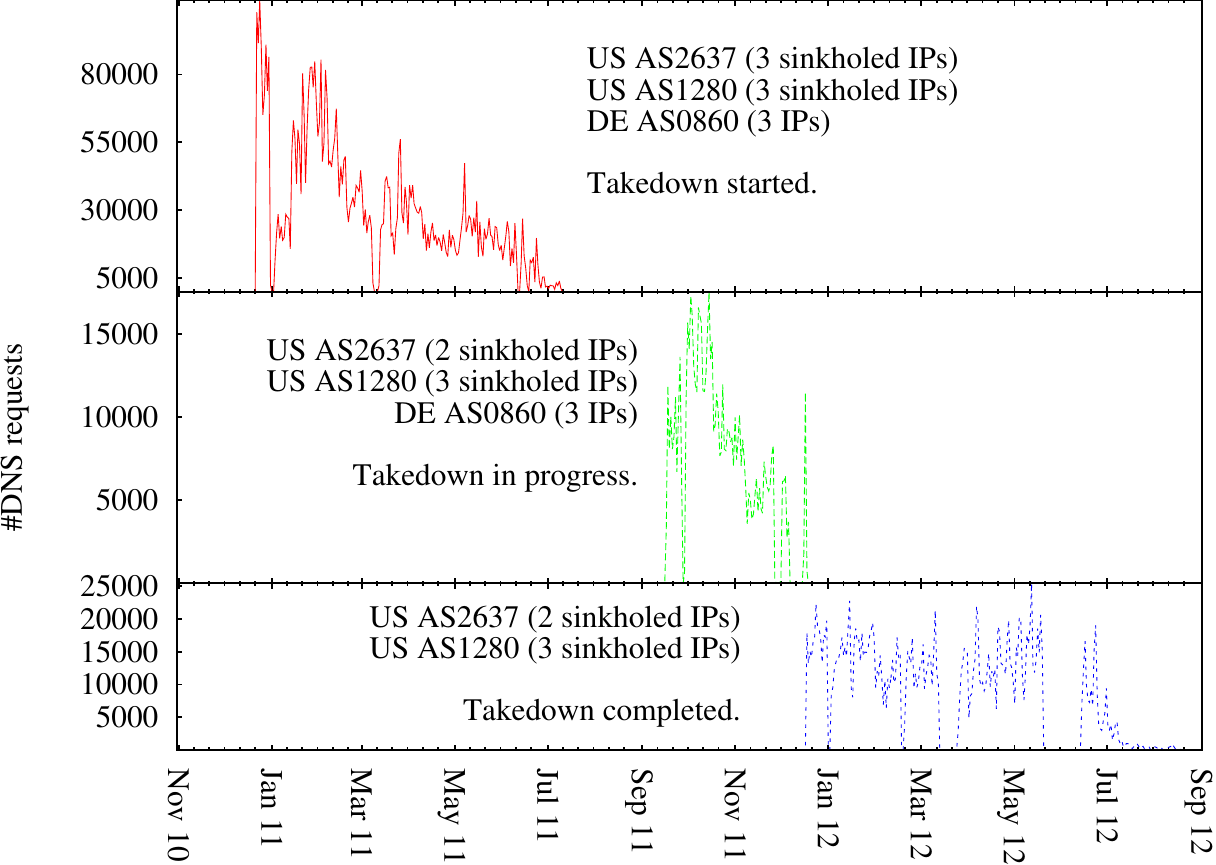}
  \caption{\textbf{Conficker}: Evolution that resembles a C\&C
    takedown: the C\&C had 3 IPs in AS0860 and 3 sinkholed IPs in
    AS2637.}
  \label{fig:time-evolution-takedown}
\end{figure}

In a similar vein, Fig.~\ref{fig:time-evolution-takedown} shows an
evolution that we may argue being a takedown operated by security
defenders. In particular, at the beginning the botnet C\&C backend was
distributed across three ASs in two countries (United States and
Germany). Armed with the knowledge that the IPs in AS2637 and AS1280
are operated by computer security laboratories, we discover that this
``waterfall'' pattern concludes into a sinkhole. Without knowledge of
the sinkholed IPs, we could still argue that the C\&C was moved from
some ASs to some other ASs.

The aforementioned conclusions were drawn by a semi-automatic analysis
and can be interpreted and used as novel intelligence knowledge. The
labels of the DGAs produced by \thesystem were fundamental to perform
this type of analysis.

\section{Discussion}
\label{sec:discussion}
Despite the good results, \thesystem has some limitations. Previous
work leveraged NXDOMAIN responses to identify those AGDs that the
botmaster did not register yet. This allows early detection of DGA
activities, because the bots yield overwhelming amounts of NXDOMAIN
replies. Our system, instead, requires \emph{registered} AGDs to
function. Therefore, it is fed with data that takes longer collection
periods. This results in a less-responsive detection of
previously unseen DGAs. The advantage is that, differently from
previous work, we can fingerprint the DGAs and, more importantly, we
lift the observation point such that \thesystem is easier to
adopt. Indeed, we believe that not using NXDOMAIN replies represents a
strength of our work, as it makes our system \hl{profoundly different
  from previous work} in ease of deployment and testing under
less-constraining requirements.

%\tod[Lore,SteZ]{Rafforzare questo concetto - \`E sufficiente?}

The linguistic features computed on the domain names, to decide
whether they are automatically generated or not, capture the
likelihood that a given domain targets English-speaking users. Taking
into account different languages, possibly featuring totally different
sounds like Chinese or Swedish, as well as different encondings, such
as UTF8, would pose some challenges. In particular, computing
language-independent features with a multilingual dictionary would
flatten the underlying distributions, rendering the language features
less discriminant. To tackle this limitation, a possible solution
consists in inferring the linguistic target of a given domain (e.g,
via TLD analysis or whois queries) so to evaluate its randomness
according to the correct dictionary.

\hl{Future DGAs may attempt to evade our linguistic features by
  creating \emph{pronounceable} AGDs. Besides the fact that, to the
  best of our knowledge, no such DGAs exist, creating \emph{large
    amounts} of pronounceable domains is difficult: Such DGAs
    would have a narrow randomization space, which
  violates the design goals of domain flux~\cite{Stone-Gross:2009:YBM:1653662.1653738,leder2009know}.}

%\tod[All]{In the same vein, how does your system cope with DGAs
%  adapting and using more english-like domains? Its worth discussing
%  (and even better experimentally validating) how susceptible your
%  approach is to evasion.}

\section{Related Work}
\label{sec:related-work}
Botnet mitigation is a very broad topic, for which we refer the reader
to an extensive survey by~\citet{bailey2009survey}. The remainder of
this section extends our overview of the state of the art in
\S\ref{sec:motivation} and discusses how our work differentiates from
or advances existing techniques. We strive to include only work that
was presented at top venues.

\paragraph{Linguistic Features of AGDs}
In this category we include works that proposed to model AGDs by means
of their linguistic features. Although the \emph{idea} of using
linguistic features is similar to what we use in \thesystem, the
existing works are based on supervised learning and make assumptions
on how domains should be grouped before processing.

\citet{yadav2010detecting} were the first who addressed the problem of
AGDs, later published also in~\cite{yadav2012detecting}: The authors
leverage the randomization of AGD names to distinguish them from
HGDs. Linguistic features capturing the distribution of alphanumeric
characters and bi-grams are computed over domain \emph{sets}, which
are then classified as sets of AGDs or HGDs. Differently from
\thesystem, their system relies on \emph{supervised} learning, and
thus requires labeled datasets of positive and negative samples. The
work explores different strategies to group domain in sets before
feeding them to the classifier: per-second-level-domain, per-IP and
per-component. The first strategy groups the domains according to
their second-level-domain, the second strategy to the IPs they resolve
to, the third to the bipartite domain-IP graph components. Our work is
different from these approaches because we require no labeled datasets
of AGDs to be bootstrapped, thus it is able to find sets of AGDs with
no prior knowledge. Moreover, our system classifies domains one by
one, without the necessity of performing error-prone apriori grouping.

\paragraph{DGA Analysis} In this category we include works that focus
on modeling DGAs and/or AGDs as a mean to detect botnet activity or to
expand existing blacklists of malicious domains. \thesystem
differentiates from these works by the type of knowledge that it
produces and by the less-demanding requirements.

\citet{perdisci2012early} focused on domains that are malicious, in
general, from the viewpoint of the victims of attacks perpetrated
through botnets (e.g., phishing, spam, drive-by download). Instead,
\thesystem focuses on AGDs and, for this reason, it models the
features of the DNS layer between bots and C\&C servers. Moreover, the
detection method of~\cite{perdisci2012early} is based on supervised
learning, whereas \thesystem uses unsupervised techniques.

\citet{neugschwandtner2011detecting} proposed a system that detects
C\&C failover strategies with techniques based on multi-path
exploration. The system explores the behavior of malware samples
during simulated network failures. Backup C\&C servers and AGDs are so
unveiled, leading to new blacklists. The approach is very promising
toward expanding blacklists of malicious domains, although it may
produce misleading results when the malware behavior depends on
time-dependent information. Differently
from~\cite{neugschwandtner2011detecting}, \thesystem discovers new
AGDs---and other knowledge---using solely passive, recursive-level DNS
traffic and requires no malware samples to work.

\paragraph{DNS Traffic Analysis} In this category we include works
that leverage features of DNS packets, at various levels of
monitoring, as a mean to find new malicious domains. \thesystem
differentiates from these works by the type of new knowledge inferred
and by the less-demanding learning technique adopted.

\citet{bilge2011exposure} proposed \textsf{\relsize{-.5}{Exposure}}, a large-scale,
passive DNS analysis technique to detect domains associated with
malicious activities, including botnet C\&C. The technique is based on
the observation that malicious domains exhibit peculiar DNS
behaviors. 15 features, ranging from time series to TTL values-based
features, are computed and used to feed a classifier trained with
real-world labelled data. The main shortcoming is that
\textsf{\relsize{-.5}{Exposure}} needs a labeled dataset of known malicious domains
for training a supervised classifier. Instead, \thesystem uses a small
feed of malicious domains to infer \emph{novel} knowledge (i.e., to
find AGDs from generally malicious domains). Instead of training a
classifier on malicious domains, we calculate thresholds for our
filters based on \emph{benign}---or, at least,
human-generated---domains.

Systems like \textsf{\relsize{-.5}{Exposure}} and
\textsf{\relsize{-.5}{Notos}}~\cite{antonakakis2010building} rely on local recursive
DNS. Instead, \textsf{\relsize{-.5}{Kopis}}~\cite{antonakakis2011detecting} analyzes
DNS traffic collected from a global vantage point at the upper DNS
hierarchy. \textsf{\relsize{-.5}{Kopis}} introduces new features such as the
requester diversity, requester profile and resolved-IPs reputation, to
leverage the global visibility and detect malicious domains. As the
authors themselves notice, \textsf{\relsize{-.5}{Kopis}} is ineffective on AGDs,
because of their short lifespan, whereas we have showed extensively
that \thesystem can detect and, more importantly, label, previously
unknown AGDs.

\citet{bilge2012disclosure} proposed DISCLOSURE, a system that detects
C\&C communications from NetFlow data analysis. Using NetFlow data
overcomes the problems of gathering raw network traffic and of
large-scale processing. However, NetFlow poses challenges on how to
use such summarized information---which, for instance, includes no
packet payload---to tell legitimate and C\&C traffic apart. Therefore,
DISCLOSURE could be used to discover domains that are malicious only
for the fact of being involved in C\&C communication, with no
indication of the DGA, if any, behind them. Instead, \thesystem
focuses exclusively on characterizing the emerging use of DGAs in C\&C
traffic.

\paragraph{Exploiting NX domains} In this category we include works
that exploit the fact that machines (i.e., bots) infected by DGA-based
malware cause the host-level DNS servers to generate
disproportionately large numbers of NX responses. In particular,
\citet{yadav2012winning} extend~\cite{yadav2010detecting} and
introduce NXDOMAINs to speedup the detection of AGDs:
\emph{registered} AGDs are recognized because they are queried by any
given client after a series of NXDOMAIN responses. The work differs
from ours substantially, mainly because it requires DNS datasets that
include the IP addresses of the querying clients. Moreover, the
approach seems fragile on sampled datasets, which is a required step
when dealing with high-traffic networks.

\section{Conclusion}
\label{sec:conclusions}
According to our extensive evaluation, \thesystem can (1) discover
(previously unknown) DGAs by telling AGDs and HGDs apart, (2) detect
previously unknown AGDs, and (3) provide insightful intelligence
(e.g., the tracking and monitoring of DGA-based C\&C domains over
time). We improve the linguistic features proposed
in~\cite{yadav2012detecting} and combine them with other features
drawn from publicly available DNS traffic. We obtained a series of
unsupervised classifiers. A known limitation of unsupervised
classifiers is that they are opaque; in other words, they do not
provide any feedback about the decisions that they draw. We overcame
this limitation by calculating fingerprints of the domains identified
by \thesystem as belonging to a group of ``similar'' AGDs.

Contrary to the existing methods based on NX domains, our approach
does not rely on clients' IPs, it is not affected by NAT and DHCP
leases nor it requires specific deployment contexts. Indeed, \thesystem
takes recursive-level DNS traffic as input, which is abundant and easily
accessible by researchers and practitioners, and a small blacklist of
malicious domains, which is also easy to obtain.

Despite the known limitations, highlighted in
\S\ref{sec:discussion}, which provide directions for further
research, we successfully used \thesystem in real-world settings to
identify a list of suspicious domains as belonging to a live botnet
(based on Conficker.B). We believe that, in addition to the
comprehensive evaluation, this latter fact proves the practicality and
effectiveness of our approach.

\section*{Acknowledgement}
This research has been partially funded under
the EPSRC Grant Agreement EP/K033344/1.

\bibliographystyle{plainnat}
\bibliography{references}

\begin{appendices}
  \section{Supplementary Information}
  A representative excerpt of the clustering produced by \thesystem
  from our dataset.

  \begin{center}
    \scriptsize
    \begin{tabular}{|>{\ttfamily}c|}
      \multicolumn{1}{c}{Cluster 6 (Bamital)} \\
      \hline
      50e7f66b0242e579f8ed4b8b91f33d1a.co.cc \\
      bad61b6267f0e20d08154342ef09f152.co.cc \\
      62446a1af3f85b93f4eef982d07cc492.co.cc \\
      0d1a81ab5bdfac9c8c6f6dd4278d99fb.co.cc \\
      f1dad9a359ba766e9f5ec392426ddd30.co.cc \\
      295e2484bddd43bc43387950a4b5da16.co.cc \\
      501815bd2785f103d22e1becb681aa48.co.cc \\
      341af50eb475d1730bd6734c812a60a1.co.cc \\
      49b24bf574b7389bd8d5ba83baa30891.co.cc \\
      a7e3914a88e3725ddafbbf67444cd6f8.co.cc \\
      \hline
    \end{tabular}
    \\[.2cm]
    \begin{tabular}{|>{\ttfamily}c>{\ttfamily}c>{\ttfamily}c|}
      \multicolumn{3}{c}{Cluster 9 (Palevo)} \\
      \hline
      7cj1b.cn & ff88567.cn & ef44ee.cn \\
      fwjp0.cn & 0bc3p.cn & 9i230.cn \\
      3dcyp.cn & azeifko.cn & fyyxqftc.cn \\
      \multicolumn{3}{|c|}{\ttfamily hfju38djfhjdi3kd.cn} \\
      \hline
    \end{tabular} \\
    [.2cm]
    \begin{tabular}{|>{\ttfamily}c>{\ttfamily}c>{\ttfamily}c>{\ttfamily}c|}
      \multicolumn{4}{c}{Cluster 10 (Palevo)} \\
      \hline ewn.net & wyp.net & ews.net & kpk.net\\
      khz.net & uon.org & lxx.net & kxc.com \\
      yhv.com & nrl.net &&\\
      \hline
    \end{tabular}
    \\[.2cm]
    \begin{tabular}{|>{\ttfamily}c>{\ttfamily}c|}
      \multicolumn{2}{c}{Cluster 11 (Conficker)} \\
      \hline byuyy.biz & jbkxbxublgn.biz \\
      kpqzk.org & tcmsrdm.org \\
      lvzqxymji.org & fbhwgmb.info \\
      aeyyiujxs.org & psaehtmx.info \\
      vdrmgyxq.biz & mmdbby.biz \\
      \hline
    \end{tabular}
  \end{center}

\end{appendices}

\end{document}